\documentclass[fleqn,usenatbib]{mnras}

\usepackage{newtxtext,newtxmath}
\usepackage[T1]{fontenc}
\usepackage[export]{adjustbox}

\DeclareRobustCommand{\VAN}[3]{#2}
\let\VANthebibliography\thebibliography
\def\thebibliography{\DeclareRobustCommand{\VAN}[3]{##3}\VANthebibliography}

\usepackage{graphicx}	% Including figure files
\usepackage{amsmath}	% Advanced maths commands
\title[Collision-induced Magnetic Reconnection]{Filament Formation via Collision-induced Magnetic Reconnection -- Formation of a Star Cluster}

\author[S. Kong et al.]{
Shuo Kong,$^{1}$\thanks{E-mail: shuokong@arizona.edu (SK)}
David Whitworth,$^{2}$
Rowan J. Smith,$^{2}$
Erika T. Hamden,$^{1}$
\\
$^{1}$Steward Observatory, University of Arizona, Tucson, AZ 85719, USA\\
$^{2}$Jodrell Bank Centre for Astrophysics, Department of Physics and Astronomy, University of Manchester, Oxford Road, Manchester M13 9PL, UK
}

\date{Accepted XXX. Received YYY; in original form ZZZ}

\pubyear{2022}

\begin{document}
\label{firstpage}
\pagerange{\pageref{firstpage}--\pageref{lastpage}}
\maketitle

% Abstract of the paper
\begin{abstract}
A collision-induced magnetic reconnection (CMR) mechanism was recently proposed to explain the formation of a filament in the Orion A molecular cloud. In this mechanism, a collision between two clouds with antiparallel magnetic fields produces a dense filament due to the magnetic tension of the reconnected fields. The filament contains fiber-like sub-structures and is confined by a helical magnetic field. To show whether the dense filament is capable of forming stars, we use the \textsc{Arepo} code with sink particles to model star formation following the formation of the CMR-filament. First, the CMR-filament formation is confirmed with \textsc{Arepo}. Second, the filament is able to form a star cluster after it collapses along its main axis. Compared to the control model without magnetic fields, the CMR model shows two distinctive features. First, the CMR-cluster is confined to a factor of $\sim4$ smaller volume. The confinement is due to the combination of the helical field and gravity. Second, the CMR model has a factor of $\sim2$ lower star formation rate. The slower star formation is again due to the surface helical field that hinders gas inflow from larger scales. Mass is only supplied to the accreting cluster through streamers. 
\end{abstract}

% Select between one and six entries from the list of approved keywords.
% Don't make up new ones.
\begin{keywords}
magnetic fields -- magnetic reconnection -- MHD -- stars: formation -- ISM: clouds -- methods: numerical
\end{keywords}

%%%%%%%%%%%%%%%%%%%%%%%%%%%%%%%%%%%%%%%%%%%%%%%%%%

%%%%%%%%%%%%%%%%% BODY OF PAPER %%%%%%%%%%%%%%%%%%

\section{Introduction}\label{sec:intro}

Filaments are crucial to star formation in giant molecular clouds \citep{2014prpl.conf...27A}, as they contain the majority of the mass budget at large column density and contain the majority of star-forming cores in the clouds \citep{2015A&A...584A..91K,2020A&A...635A..34K}. Understanding filament formation thus becomes a crucial part of a complete picture of star formation \citep{2019A&A...623A.142S}. Previously, ideas of filament formation include turbulent shocks \citep[e.g.,][]{2001ApJ...553..227P}, sheet fragmentation \citep[e.g.,][]{2009ApJ...700.1609M}, magnetic-field channeling \citep[e.g.,][]{2019MNRAS.485.4509L}, and Galactic dynamics \citep[e.g.,][]{2020MNRAS.492.1594S}. A summary of filament formation mechanisms can be found in the latest review in \citet{2022arXiv220309562H}. 

Recently, \citet[][hereafter K21]{2021ApJ...906...80K} demonstrated a new mechanism of filament formation via collision-induced magnetic reconnection (CMR). The study was motivated by the special morphology of the sub-structures of the Stick filament that resemble those created by magnetic reconnection. Given the fact that Orion A is between a large-scale magnetic field reversal \citep{1997ApJS..111..245H} and the position-velocity (PV) diagram shows two velocity components, K21 proposed the scenario in which two clumps collide with antiparallel magnetic fields. The model successfully reproduced observational features of the Stick filament, including the morphology (ring/fork-like structures), the density probability distribution function (PDF), the line channel maps, and the PV diagrams. Moreover, the model results gave an alternative explanation to the findings in \citet{2017ApJ...846..144K} that cores in Orion A were mostly pressure-confined. The natural result of the helical field around the filament exerts a surface magnetic pressure on the filament, confining the filament and the cores. For the first time, the CMR model provides a complete picture of structure formation in Orion A that self-consistantly incorporates the 25-year mystery of the reversed magnetic field.

\begin{figure*}
\centering
\includegraphics[width=0.49\textwidth]{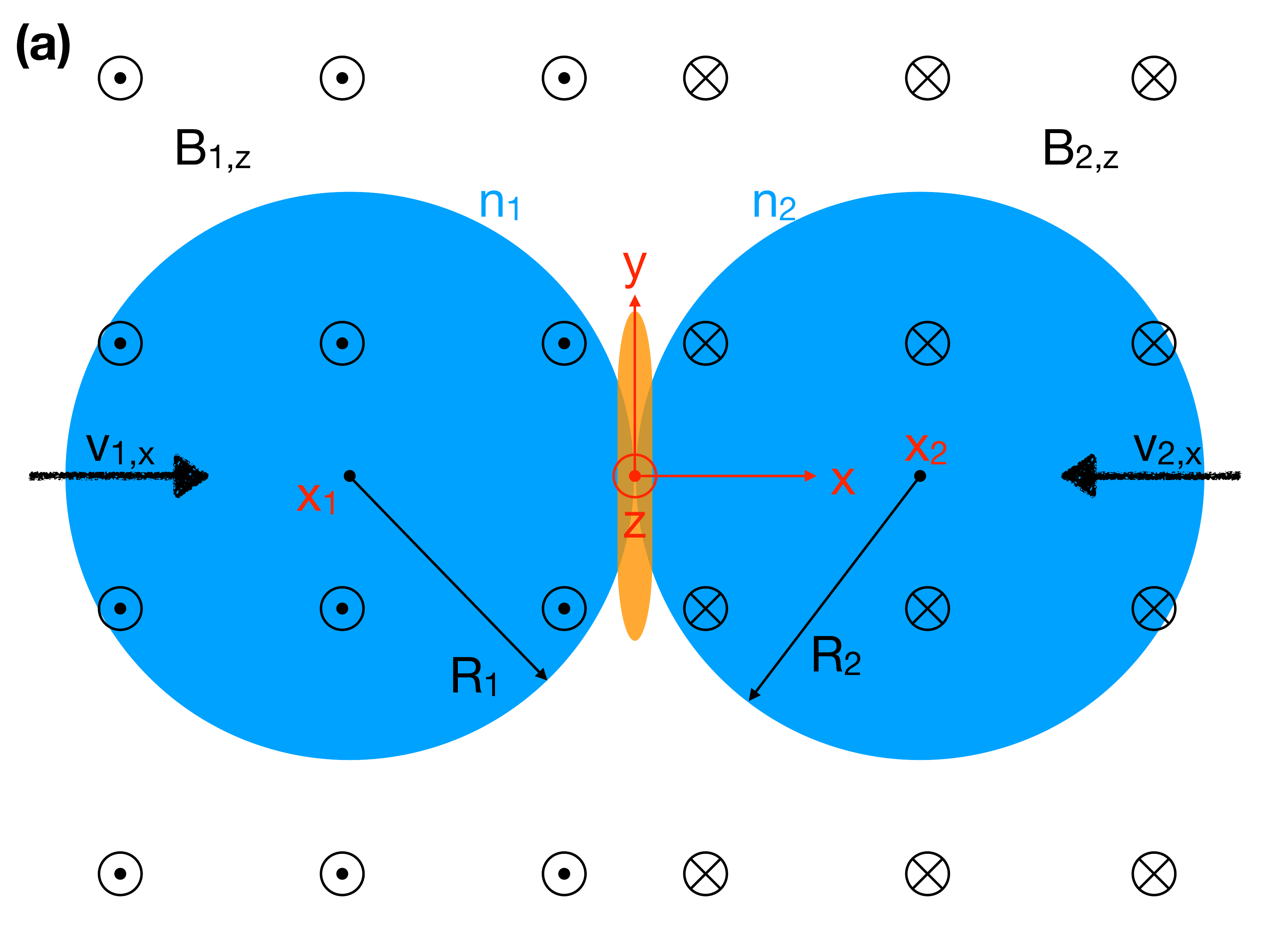}
\includegraphics[width=0.49\textwidth]{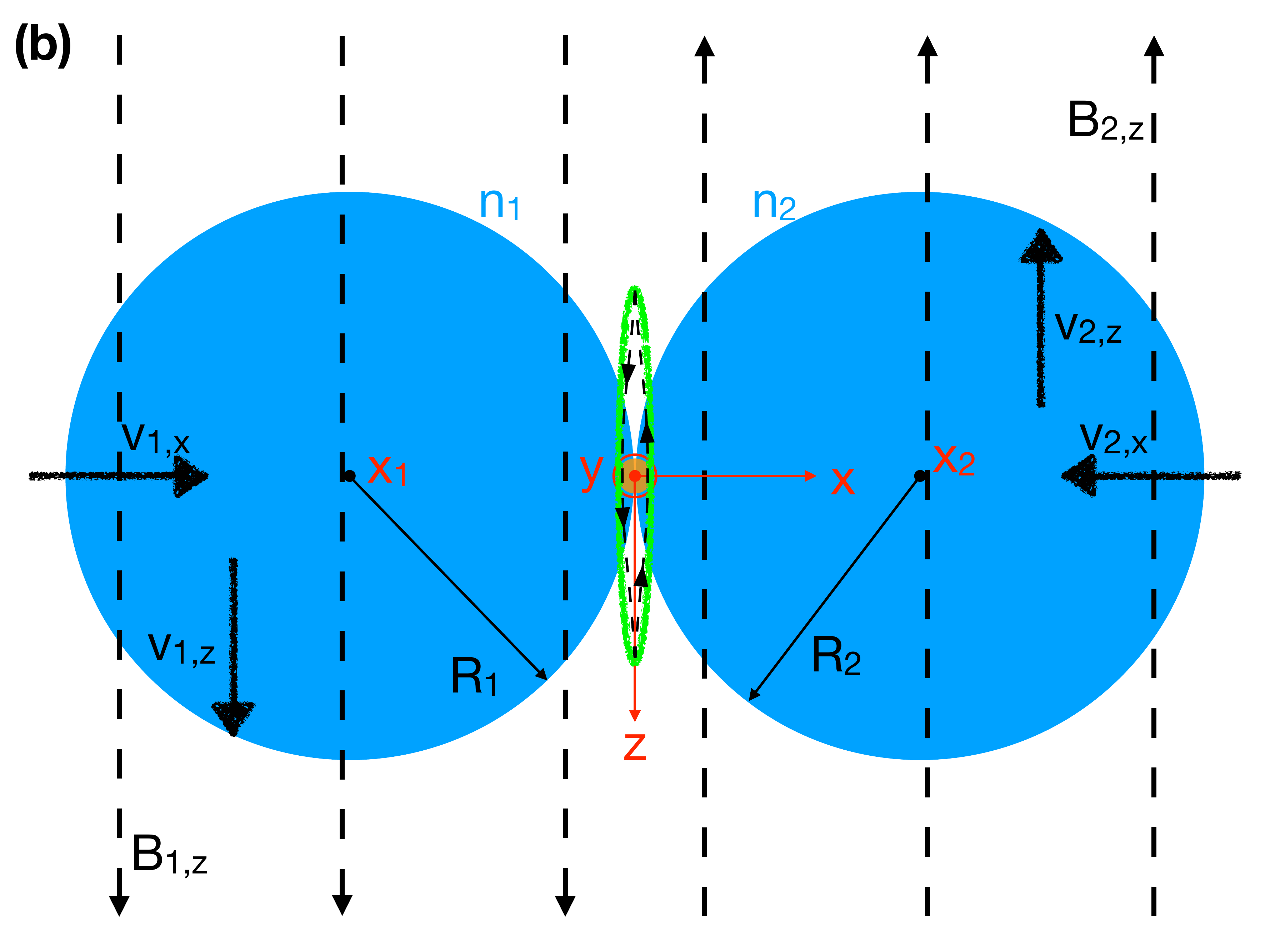}
\caption{
An illustration of CMR in two viewing angles. {\bf (a):} A view in the x-y plane. The Cartesian coordinate system (red) centers at the collision point. The x-axis points rightward and the y-axis points to the top. The z-axis points toward us as indicated by the red circle-point. The clouds have colliding velocities $v_{\rm 1,x}$ and $v_{\rm 2,x}$, respectively. The magnetic field points toward us (marked as black circle-points) for $x<0$ and away from us (marked as black circle-crosses) for $x>0$. After collision, the filament (orange) forms along the y-axis. {\bf (b):} A view in the z-x projection. In this view, the magnetic field is parallel to the plane of the sky. The y-axis points toward us as indicated by the red circle-point. After collision, the filament (orange) forms along the y-axis which points toward us. The green ellipse marks the location of the compression pancake if no magnetic fields. With antiparallel fields and CMR, the field reconnects at two tips of the pancake and forms a loop (black dashed arrow curve) around the pancake. Due to the magnetic tension force, the pancake is squeezed into the central axis (y-axis) becoming a filament.}
\label{fig:cmr}
\end{figure*}

Figure \ref{fig:cmr} illustrates the CMR filament formation. In panel (a), we view the process from the side of the filament. Two clouds move along the x-axis and collide at the origin. On the left side of the y-z plane, the magnetic field points toward us. On the other side, the field points away from us. After collision, the reversed field reconnects in the z-x plane and forms field loops that pull the compression pancake into the central axis (y-axis in our setup). The pulling is due to the magnetic tension the field loop exerts on the gas. As a result, a filamentary structure forms along the y-axis. In panel (b), we view the process in the z-x plane. In this projection, we are looking at the filament cross-section at the origin. The green ellipse represents the compression pancake and the black dashed arrow curve around the pancake denotes the reconnected field loop. The loop has a strong magnetic tension that pulls the dense gas in the pancake to the origin in each z-x plane. As a result, the filament (orange cross-section) forms along the y-axis. Essentially, the filament forms along the field symmetry axis that crosses the collision point. 

While K21 outlined the skeleton of the theory, more follow-up studies are needed to further understand the physical process. Among the unknowns about CMR, the most urgent one is whether a CMR-filament can produce stars. While K21 showed that CMR can quickly make dense gas with $n_{\rm H_2}\sim10^5~{\rm cm}^{-3}$, it was not obvious that the dense gas would eventually collapse and form stars instead of being transient in the interstellar medium. 

In this paper, we aim to confirm star formation within CMR-filaments, and compare it with star formation in other types of conditions. We will see how CMR star formation differs from other star formation pathways. In the following, we introduce the numerical method in \S\ref{sec:method}. Then, in \S\ref{sec:ic}, we describe the initial conditions for our fiducial model. In \S\ref{sec:results}, we present results from the simulations. Finally, we summarize and conclude in \S\ref{sec:conclu}.

\section{Method}\label{sec:method}

We use a modified version of the \textsc{Arepo} code \citep{Springel10} to model the formation of the filament. In particular, we simulate the compressible and inviscid magnetohydrodynamics (MHD). The code adopts the finite-volume method on an unstructured Voronoi grid that is dynamically created from mesh generating points that move according to the local velocity of the fluid. The target mass contained within each cell can be arbitrarily selected by the user, meaning that the spatial resolution of \textsc{Arepo} varies according to the local gas density. In our simulations we set a default target mass for each cell of $3.6\times10^{-4}$ M$_\odot$, however we also require that the Jeans scale be resolved by a minimum of 16 cells as to avoid artificial fragmentation \citep{Truelove97} and ensure many cells span the width of the filament.

The implementation of magnetic fields in \textsc{Arepo} was described in \citet{Pakmor11} and uses a HLLD Riemann solver and Dedner divergence cleaning. Gravity is included using a tree-based approach improved and modified for \textsc{Arepo} from \textsc{Gadget-2} \citep{Springel05}. When calculating the gravitational forces we do not use periodic boundaries.

We use a custom implementation of chemistry whose development is described in \citet{Smith14a,Clark19}. The gas chemistry is based off the network of \citet{Gong17} and was first implemented in \textsc{Arepo} in \citet{Clark19}. The \citet{Gong17} network was designed to accurately reproduce the CO abundances in low density regions using a 1D equilibrium model, but in high density regions may over-produce atomic carbon. Our implementation is a non-equilibrium, time-dependent 3D version of the above that contains several additional reactions that are unimportant in PDR conditions but that make the network more robust when dealing with hot, shocked gas. Full details of these modifications can be found in Hunter et al (in prep.).

Heating and cooling of the gas is computed simultaneously with the chemical evolution using the cooling function described in \citet{Clark19}. To do this accurately it is important to calculate the local shielding from dust and H$_2$ self shielding with respect to the Interstellar Radiation Field (ISRF). We calculate this using the \textsc{TreeCol} algorithm that \citet{Clark12b} first implemented in \textsc{Arepo}. The background radiation is assumed to be constant at the level calculated by \citet{Draine78} and enters uniformly through the edges of the box. Cosmic ray ionisation is assumed to occur at a rate of $3 \times 10^{-17}$ s$^{-1}$.

Star formation is modelled within the code using sink particles \citep{Bate95,Greif11}. Above number densities of $n_{\rm H_2}\sim10^8~{\rm cm}^{-3}$, we check whether the densest cell in the deepest potential well and its neighbours satisfy the following three conditions: (1) the cells are gravitationally bound, (2) they are collapsing, and (3) the divergence of the accelerations is less than zero, so the particles will not re-expand (see also \citealt{Federrath10a}). If all these conditions are satisfied the cell and its neighbours are replaced with a sink particle, which interacts with the gas cells purely through gravitational forces. Additional material can be accreted by the sink particles from neighbouring cells. This occurs via skimming mass above this density threshold if the adjacent cells move within an accretion radius about three times the Jeans scale (0.0018 pc) and are gravitationally bound to it. In our current study we focus on the early stages of star formation at the core fragmentation phase, and therefore we neglect any radiative feedback from the sinks, which would play a role later in the evolution.

We adopt the same unit system as K21. Specifically, the code unit for mass density is $3.84\times10^{-21}$ g cm$^{-3}$ ($n_{\rm H_2}$=840 cm$^{-3}$, assuming a mean molecular mass per H$_2$ of $\mu_{\rm H_2}=2.8 m_\textrm{H}$). The code unit for time is 2.0 Myr. The code unit for length scale is 1.0 pc. The code unit for velocity is 0.51 km s$^{-1}$. With these settings, the gravitational constant is $G=1$, and the magnetic field unit is 3.1 $\mu$G.

\section{Initial Conditions}\label{sec:ic}

\begin{table}
% \normalsize
% \centering
\caption{Model Parameters.}\label{tab:ic}
\begin{tabular}{cccc}
\hline
Parameters & \mbox{MRCOLA} & \mbox{COLA\_sameB} & \mbox{COLA\_noB}\\
\hline
$L$ & 8 pc & 8 pc & 8 pc \\
$T_{\rm dust}$ & 15 K & 15 K & 15 K \\
$T_{\rm gas}$ & 15 K & 15 K & 15 K \\
$\zeta$ & $3.0\times10^{-17}$ s$^{-1}$ & $3.0\times10^{-17}$ s$^{-1}$ & $3.0\times10^{-17}$ s$^{-1}$ \\
$G$ & 1.7$G_0$ & 1.7$G_0$ & 1.7$G_0$ \\
DGR & $7.09\times10^{-3}$ & $7.09\times10^{-3}$ & $7.09\times10^{-3}$ \\
$n_{\rm amb}$ & 42 cm$^{-3}$ & 42 cm$^{-3}$ & 42 cm$^{-3}$ \\
\hline
$n_1$ & 420 cm$^{-3}$ & 420 cm$^{-3}$ & 420 cm$^{-3}$ \\
$x_1$ & -0.9 pc & -0.9 pc & -0.9 pc \\
$R_1$ & 0.9 pc & 0.9 pc & 0.9 pc \\
$v_{\rm 1,x}$ & 1.0 km s$^{-1}$ & 1.0 km s$^{-1}$ & 1.0 km s$^{-1}$ \\
$v_{\rm 1,z}$ & 0.25 km s$^{-1}$ & 0.25 km s$^{-1}$ & 0.25 km s$^{-1}$ \\
$B_{\rm 1,z}$ & 10 $\mu$G & 10 $\mu$G & 0 \\
\hline
$n_2$ & 420 cm$^{-3}$ & 420 cm$^{-3}$ & 420 cm$^{-3}$ \\
$x_2$ & 0.9 pc & 0.9 pc & 0.9 pc \\
$R_2$ & 0.9 pc & 0.9 pc & 0.9 pc \\
$v_{\rm 2,x}$ & -1.0 km s$^{-1}$ & -1.0 km s$^{-1}$ & -1.0 km s$^{-1}$ \\
$v_{\rm 2,z}$ & -0.25 km s$^{-1}$ & -0.25 km s$^{-1}$ & -0.25 km s$^{-1}$ \\
$B_{\rm 2,z}$ & -10 $\mu$G & 10 $\mu$G & 0 \\
% \hline
% $Z$ & 0.1 & metalicity \\
% $[C/H]$ & 0 & C abundance \\
% $[N/H]$ & 0 & N abundance \\
% $[O/H]$ & 0 & O abundance \\
\hline
\end{tabular}\\
{$L$ is the domain size. $T_{\rm dust}$ is the initial dust temperature. $T_{\rm gas}$ is the initial gas temperature. $\zeta$ is the cosmic-ray ionization rate. $G$ is the ISRF in unit of Habing field $G_0$. DGR is the dust-to-gas mass ratio. $n_{\rm amb}$ is the ambient H$_2$ number density. $n_1$ is the Cloud1 H$_2$ number density. $x_1$ is the Cloud1 location. $R_1$ is the Cloud1 radius. $v_{\rm 1,x}$ is the Cloud1 collision velocity. $v_{\rm 1,z}$ is the Cloud1 shear velocity. $B_{\rm 1,z}$ is the Cloud1 B-field. $n_2$ is the Cloud2 H$_2$ number density. $x_2$ is the Cloud2 location. $R_2$ is the Cloud2 radius. $v_{\rm 2,x}$ is the Cloud2 collision velocity. $v_{\rm 2,z}$ is the Cloud2 shear velocity. $B_{\rm 2,z}$ is the Cloud2 B-field. See Figure \ref{fig:cmr} for illustration.}
\end{table}

The setup for the fiducial model follows the K21 fiducial model (see K21 Figure 8), but with several additional parameters. Following the nomenclature in K21, we name our fiducial model \mbox{MRCOLA} (MRCOL+\textsc{Arepo}). As shown in Figure \ref{fig:cmr}, Cloud1 has density $n_1$, radius $R_1$, colliding velocity $v_{\rm 1,x}$ (positive x), shear velocity $v_{\rm 1,z}$ (positive z), magnetic field $B_{\rm 1,z}$ (positive z); Cloud2 has $n_2$, radius $R_2$, colliding velocity $v_{\rm 2,x}$ (negative x), shear velocity $v_{\rm 2,z}$ (negative z), magnetic field $B_{\rm 2,z}$ (negative z). Table \ref{tab:ic} lists the values for these parameters. They are the same as those in K21. With our adopted reference cell mass, the equivalent cell size in the cloud is 0.014 pc before refinement, which is about twice the cell size in K21. 

To avoid artifacts due to the periodic boundary condition, the computation domain is enlarged to 8 pc in each dimension, which is twice the size of the computation domain in K21. This is because density waves due to the colliding gas could propagate through the boundaries and impact the dynamical evolution of the filament and the sink formation. The new setup has more padding area between the clouds and the boundaries, so the boundary waves do not affect the central filament before $t=3$ Myr (the ending time).
 
We follow K21 to adopt an initial dust and gas temperature of 15 K. In turn, we assume a fully molecular composition for the system simply due to the low temperature. We include the standard ISRF of 1.7$G_0$ that illuminates the computation domain from all directions. Here $G_0$ is the Habing field. The cosmic-ray ionization rate is fixed at $3.0\times10^{-17}~{\rm s}^{-1}$. A standard dust-to-gas mass ratio of 1/141 is adopted. Table \ref{tab:ic} summarizes the parameters.

\section{Results and Analysis}\label{sec:results}

\subsection{Fiducial model}\label{subsec:fiducial}

\begin{figure*}
\centering
\includegraphics[width=\textwidth]{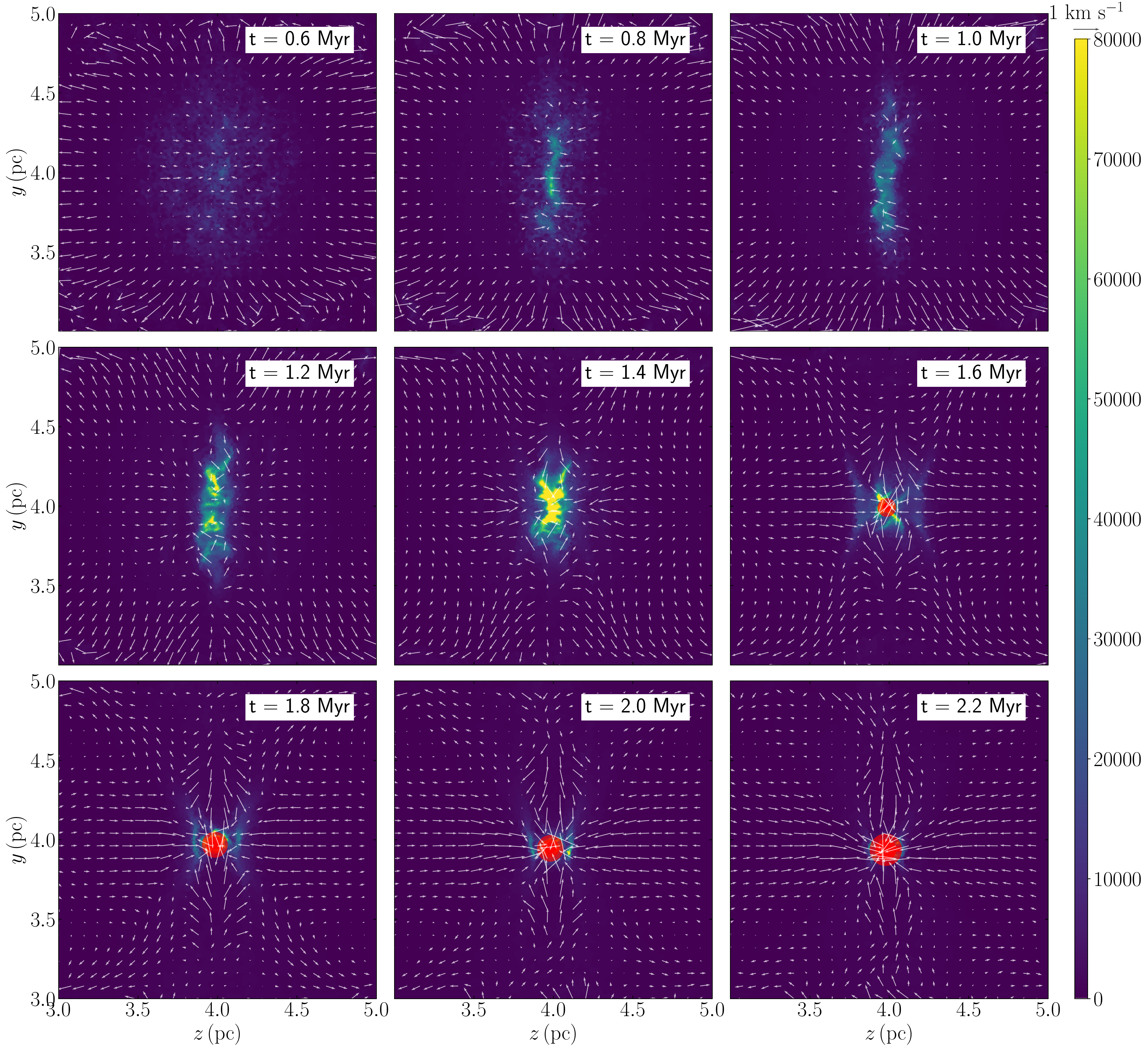}
\caption{
Density slice plots for the collision midplane (x=4 pc) as a function of time for \mbox{MRCOLA}. The color plot is in unit of $n_{\rm H_2}$ (cm$^{-3}$). The time step is shown at the upper right. The white arrows show the velocity vectors in the plane. Their lengths are proportional to the magnitudes. The red circles show the sink locations. Their sizes are proportional to the sink masses.}\label{fig:mrcolavxlin}
\end{figure*}

Figure \ref{fig:mrcolavxlin} shows density slice plots for the x=4 pc plane as a function of simulation time (upper right). The color background shows the density field. We use a linear color scale to highlight the clumpy structures. The white vectors show the velocity field. Here we only include snapshots from t=0.6 Myr to t=2.2 Myr. We also zoom in to the central 2 pc region to focus on the filament. A more complete view of the domain is shown in Appendix \S\ref{app:fiducial}.

The slice is the collision midplane where the compression pancake forms. The pancake is the dense structure in the central region at t=0.6 Myr (more prominent at t$<$0.6 Myr in Figure \ref{fig:mrcolavx}). It pushes gas outwards at its periphery, so we see the radial velocity vectors at the boundaries. In the central 1 pc region, however, the velocity vectors point toward the z=4 pc axis. The inward velocity is caused by the magnetic reconnection. The reconnected field pulls the gas toward the central axis, as shown in \S\ref{sec:intro}. Through t=0.8 Myr, the inward velocity persists and more material continues to be pulled to the central axis where the filament forms (also see Figure \ref{fig:mrcolavy}). Note the filament has always been clumpy.

\begin{figure*}
\centering
\includegraphics[width=\textwidth]{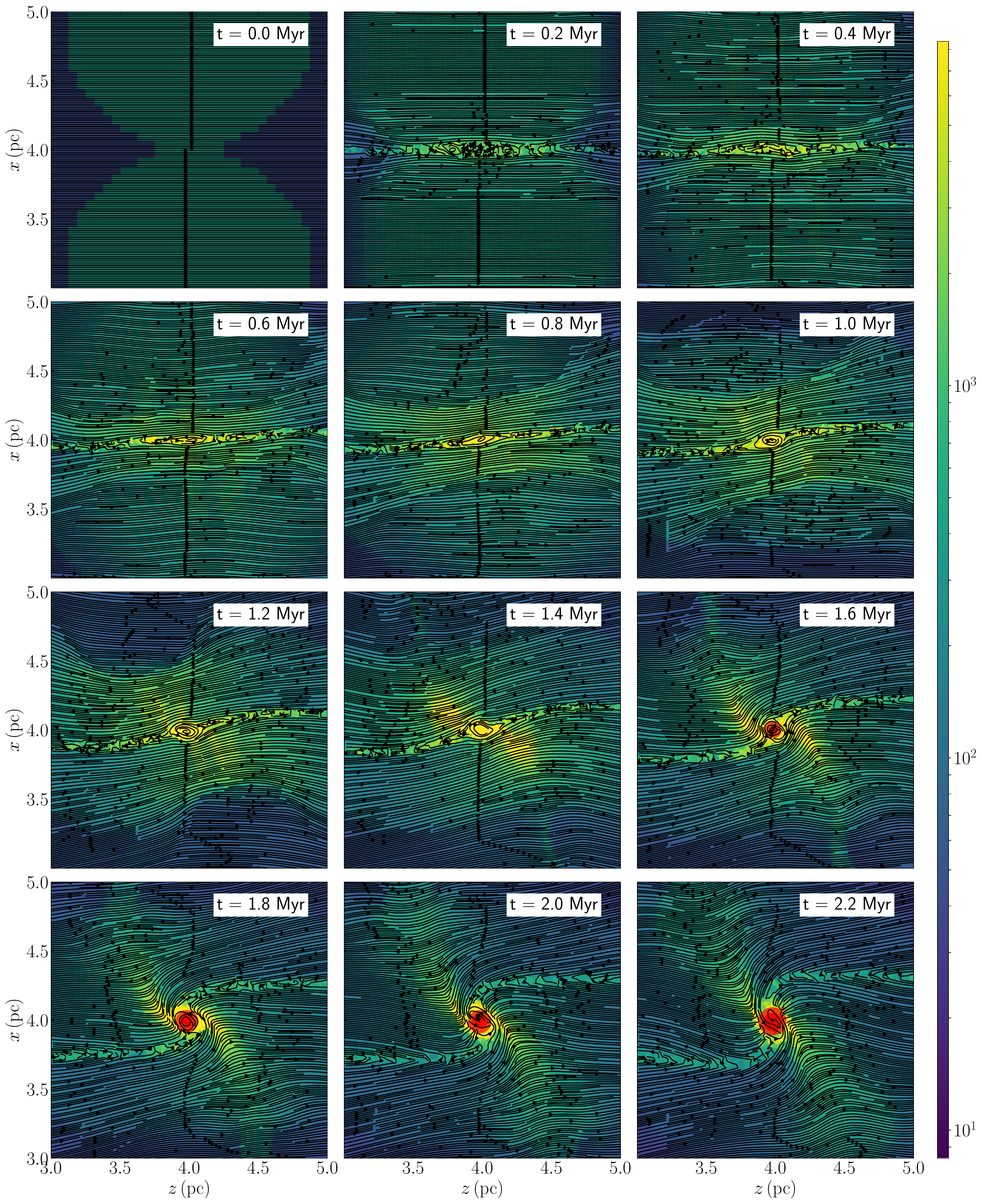}
\caption{
Density slice plots for the y=4 pc plane as a function of time. The color plot is in unit of $n_{\rm H_2}$ (cm$^{-3}$). The time step is shown at the upper right. We are viewing the filament cross-section at (x,z)=(4 pc,4 pc). Magnetic fields are shown as arrow stream lines in the plots.}\label{fig:mrcolaybb}
\end{figure*}

Figure \ref{fig:mrcolaybb} shows the CMR phenomenon in a different angle. Here we show the y=4 pc slice at different time steps. Magnetic field lines are overlaid on the density slice plots. In the center, we see the cross section of the filament, which is wrapped by circular fields. The circular fields are results of magnetic reconnection at the two ends of the compression pancake. The reconnection creates field loops which enclose the pancake. The magnetic tension squeeze the pancake to form the filament in the center. The process is the same as that shown in K21. Therefore, the filament formation mechanism through CMR is confirmed with \textsc{Arepo}.

In Figure \ref{fig:mrcolavxlin}, the filament continues to become denser through t=1.2 Myr. At t=1.4 Myr, the filament starts to collapse along its main axis, which is also indicated by the longitudinal velocity vectors in the filament. Meanwhile, gas in the vicinity of the filament shows converging velocity vectors, especially in the horizontal directions. The convergence indicates that the filament gravity dominates the central region and the region begins a global collapse. 

In fact, the collapsing gas spirals into the filament. Figure \ref{fig:mrcolavy} shows the gas kinematics better in the y=4 pc plane. Here, the x=4 pc line corresponds to the slice of Figure \ref{fig:mrcolavxlin}. At t$\gtrsim$1.2 Myr, we can see the gas around the central filament (cross-section) spiraling toward the filament. Note, the horizontal inflow velocity in Figure \ref{fig:mrcolavxlin} t=1.2 Myr panel is not the gas flow along the field reversal plane, which has a spiral shape in Figure \ref{fig:mrcolavy}. Due to the reconnected field, dense gas along the field-reversal plane continues to be dragged into the central filament, which is shown in the t=1.2 Myr panel in Figure \ref{fig:mrcolaybb}. However, this field-reversal plane is not captured in the x=4 pc slice plot in Figure \ref{fig:mrcolavxlin}. The horizontal inflowing gas in Figure \ref{fig:mrcolavxlin} is indeed due to gravity.

Also shown in the t=1.2 Myr panel of Figure \ref{fig:mrcolavy} are multiple striations perpendicular to the field reversal plane. They are also perpendicular to the incoming spiral velocity and the magnetic field (Figure \ref{fig:mrcolaybb}). If we look at the x=4 pc plane which is shown in Figure \ref{fig:mrcolavx}, there are multiple vertical striations parallel to the central filament. Now we again look at Figure \ref{fig:mrcolavy}, we realize that the striations are actually dense sheets perpendicular to the magnetic field. The spiral-in gas moves along the field lines and accumulates in sheets, similar to what was seen in previous studies, e.g., \citet{2007MNRAS.382...73T}. Later, these sheets merge into a spiral structure (nearly perpendicular to the field-reversal) to be accreted by the filament.

In Figure \ref{fig:mrcolavxlin}, at t=1.6 Myr, the filament almost shrinks to become a dense core while the collapse continues along the horizontal and vertical directions. Some gas moves away from the region through an X-shaped outflow (not a protostellar outflow). Until now, no sinks form. So at least in the fiducial model \mbox{MRCOLA}, the clumpy dense gas initially in the filament is not able to form stars. In fact, the formation of the clumpy gas is different from other filament models in which dense clumps form due to fragmentation of a critical filament. Here, the dense gas is moved and bound to the central axis piece by piece. The gas is already clumpy during the transportation \citep[cf.][for a similar but not identical scenario in which sub-filaments merge into a single large structure]{2014MNRAS.445.2900S}. The gas clumps constitute the filament. It is almost a reverse process of fragmentation. Essentially, the filament morphology is determined by the dynamics due to the magnetic field. 

\subsection{Cluster Formation}\label{subsec:sink}

\begin{figure*}
\centering
\includegraphics[width=\textwidth]{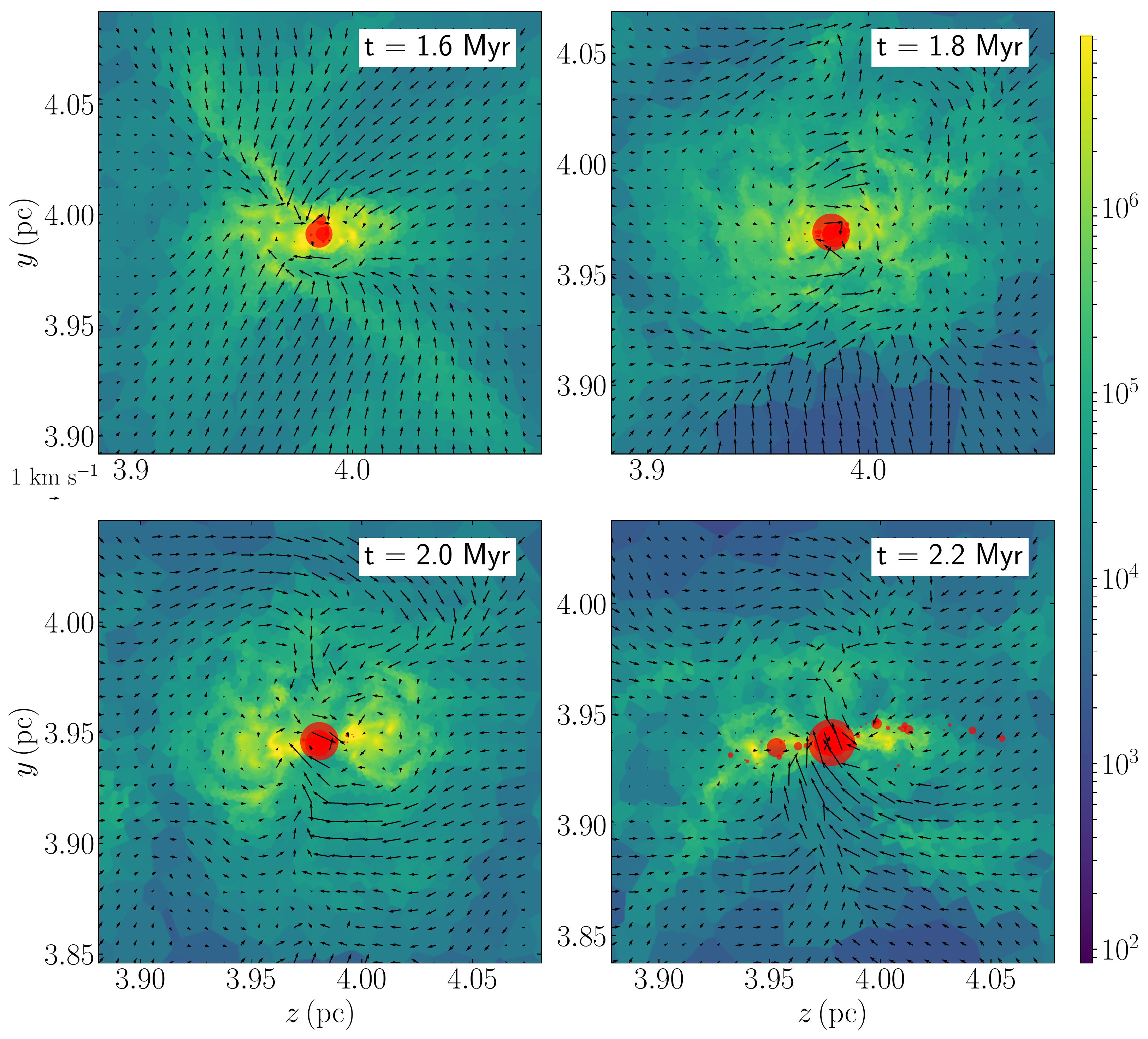}
\caption{
Zoom-in view of the density slice plot for the x=4 pc plane in the fiducial model. The color plot is in unit of $n_{\rm H_2}$ (cm$^{-3}$). Each slice plot centers at the mass-weighted cluster center. The arrows show the velocity vectors in the plane. Their sizes are proportional to the magnitudes. The red semi-opaque circles show the sink location. Their sizes are proportional to the sink mass.}\label{fig:zvx}
\end{figure*}

\begin{figure*}
\centering
\includegraphics[width=\textwidth]{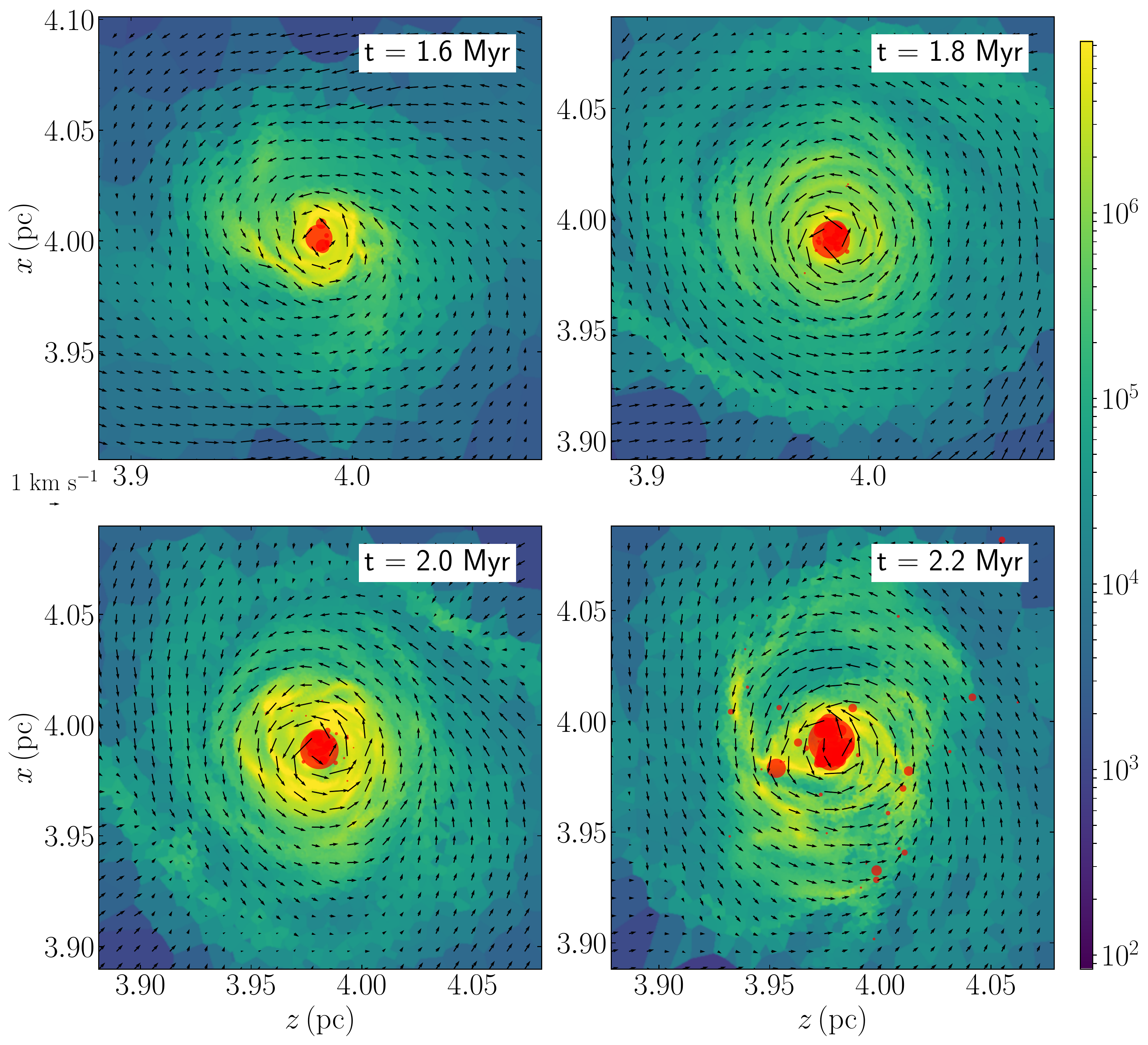}
\caption{
Zoom-in view of the density slice plot for the y=4 pc plane in the fiducial model. The format is the same as Figure \ref{fig:zvx}.} \label{fig:zvy}
\end{figure*}

\begin{figure*}
\centering
\includegraphics[width=\textwidth]{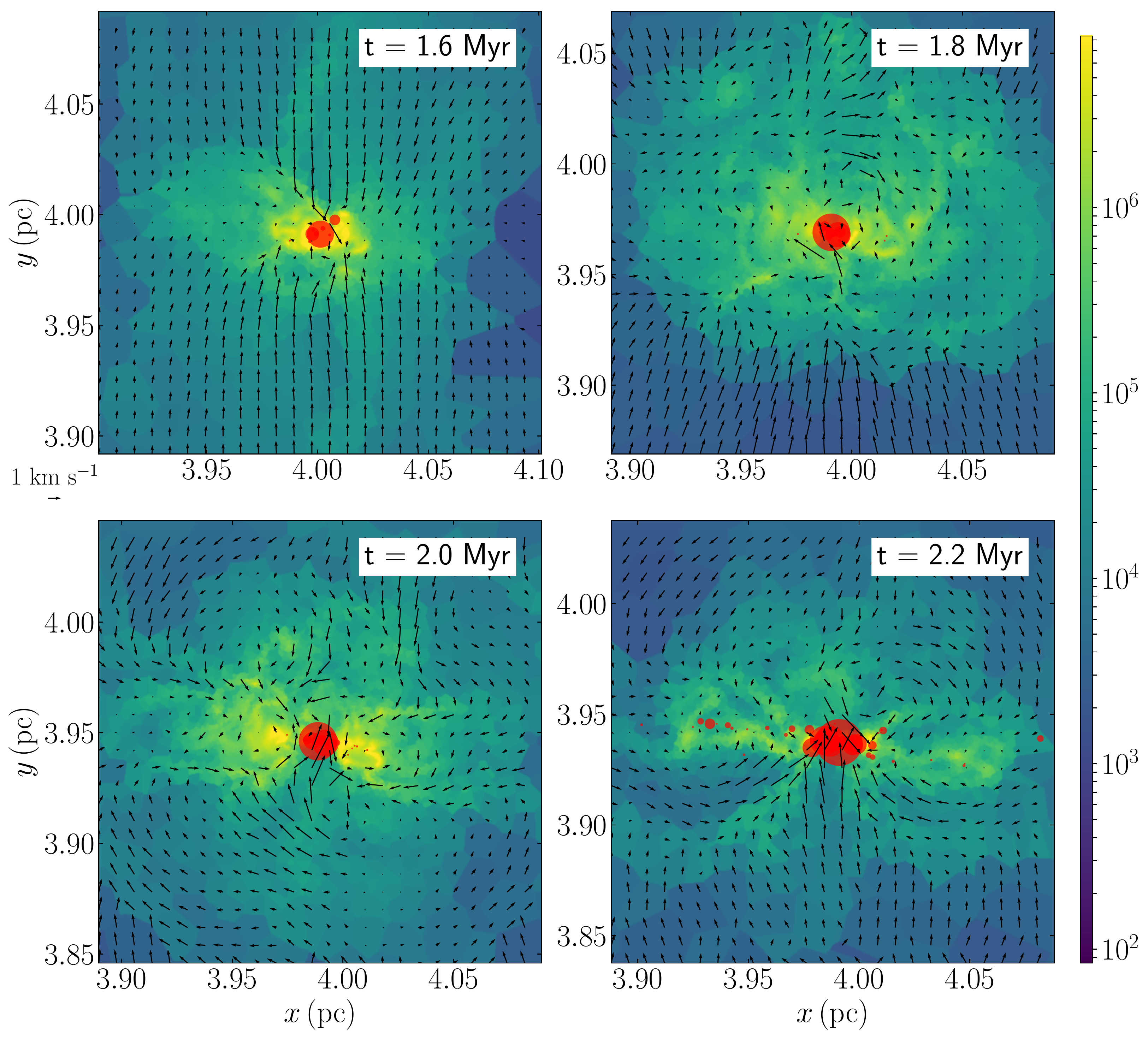}
\caption{
Zoom-in view of the density slice plot for the z=4 pc plane in the fiducial model. The format is the same as Figure \ref{fig:zvx}.}\label{fig:zvz}
\end{figure*}

Once the filament starts to collapse along its main axis, dense gas accumulates in the central region, which we term the dense core \citep[not necessarily the dense core in observations, e.g.,][]{2018ApJ...855L..25K,2019ApJ...873...31K,2021ApJ...912..156K}. Soon after t=1.4 Myr, the first sink forms in the core. By t=1.6 Myr, 9 sinks are present in the core, as indicated by the red circles. The sink formation indicates that the CMR mechanism is capable of forming stars, which answers the opening question in \S\ref{sec:intro}. Not only does CMR form stars, it is capable of producing a cluster (see below). However, the star formation does not happen during the initial filament phase, but happens after the filament collapses into a central dense core.

In Figure \ref{fig:mrcolavxlin}, the collapse continues from t=1.8 Myr to t=2.2 Myr. The sinks grow more massive by accreting the inflowing gas. The most massive sink in the t=2.2 Myr panel is 8.1 M$_\odot$. Meanwhile, dense gas continues to flow toward the cluster forming region, as indicated by the converging velocity vectors, feeding the mini cluster. The central dense core and the star cluster grow together, showing a concurrent, dynamical star cluster formation picture. By the time of 2.2 Myr, some sinks should probably have protostars and their feedback should change the subsequent fragmentation and accretion. Since we do not include the feedback, we do not continue the simulation further.

Combining all the analyses above, we can see that overall the CMR star formation (CMR-SF) is a two-phase process, at least in the specific model of \mbox{MRCOLA}. First, a dense, clumpy filament forms due to magnetic tension. Second, the filament collapses and forms a dense core in which a star cluster emerges. 

To better show the cluster structure and how the collapsing gas feeds the cluster growth, we zoom in to the central 0.2 pc region and show the slice plots. Figures \ref{fig:zvx}, \ref{fig:zvy}, \ref{fig:zvz} show the zoom-in view of constant x, y, z planes, respectively. Each slice plot centers at the mass-weighted cluster center. The red filled circles show the sinks.

From the three figures we can see that the dense gas around the cluster is chaotic. Figure \ref{fig:zvy} shows spiral gas structures and velocities, consistent with our interpretation on the large scale in \S\ref{subsec:fiducial}. As we discussed, the spiral structure originates from the initial shear velocity. However, it does not develop into a flat disk, as we can see in Figures \ref{fig:zvx} and \ref{fig:zvz}. Perhaps a flat structure is visible at t=2.2 Myr. But more often, the dense gas is in disorder. Sometimes, there are gas streamers that show coherent inflow velocities toward the cluster. They are the main source of mass supply that feeds into the accreting cluster. In contrast, the same cloud-cloud collision without magnetic fields develops a flat disk starting from 1.0 Myr (see \S\ref{subsec:control}).

As shown in Figures \ref{fig:zvx} and \ref{fig:zvz}, the cluster is generally distributed in a constant-y plane, more so for $t=2.2$ Myr. Figure \ref{fig:zvy} shows that the cluster rotates in the constant-y plane, following the rotation of the dense gas. The angular momentum of the cluster, which it inherits from the gas, makes the cluster settle in a stellar disk. The reason the cluster is not as chaotic as the dense gas is because the sinks only interact with the gas through gravity, i.e., they do not feel the gas pressure or the magnetic field.

There is only one cluster in the computation domain and it is highly concentrated within a diameter $\lesssim$ 0.05 pc ($\sim10^4$ AU). The cluster concentration is largely due to the dense gas concentration. Again in Figure \ref{fig:zvy}, we can see that the size of the densest gas is also about 0.05 pc, just enclosing the cluster. Here, the gas density reaches $\gtrsim10^7$ cm$^{-3}$. Outside the cluster, the gas streamers/spirals connect the system to the larger collapsing region. 

Within the cluster, we can see that the most massive members tend to stay at the center. This apparent mass segregation is more prominent from t=2.0 Myr to t=2.2 Myr. The segregation is not surprising because those stars closer to the collapse center form earlier and have the advantage of accreting denser gas, which is similar to the idea of ``Competitive Accretion'' \citep{2001MNRAS.323..785B,2006MNRAS.370..488B}, where the mass segregation is not the result of initial condition but a natural result of accretion at different locations. 

\begin{figure}
\centering
\includegraphics[width=\columnwidth]{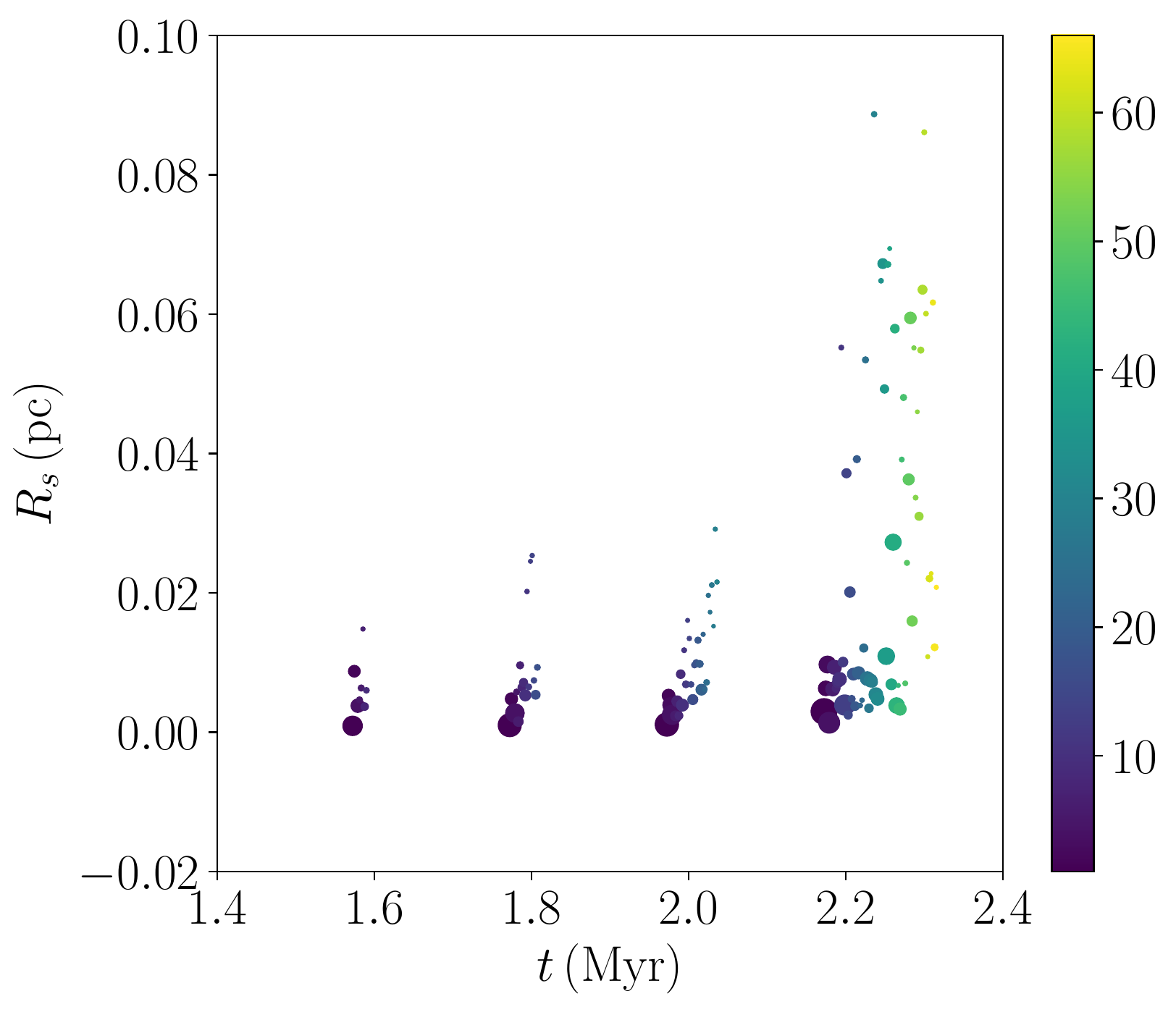}
\caption{
Distance from a sink to the mass-weighted center of the cluster $R_s$ as a function of time $t$. The color shows the sink ID. Larger IDs indicate later formation time. The size of the circle represents the sink mass. The normalization is different from previous Figures. To reduce overlap, we spread the circles along the horizontal axis, while the valid time steps only include 1.6, 1.8, 2.0, and 2.2 Myr.}\label{fig:tr}
\end{figure}

In Figure \ref{fig:tr}, we show sink locations as a function of time. $R_s$ is defined as the distance from the sink to the mass-weighted cluster center. Darker colors indicate sinks formed earlier (marked with increasing integers). We can see that the most massive sinks at t=2.2 Myr are those formed the earliest. They also stay near the cluster center all the time. Those formed at larger distances do not grow as massive as those near the center. Meanwhile, new members (lighter colors) emerge at different radii. Those near the center will likely grow faster than those farther away.

\subsection{Control models}\label{subsec:control}

\begin{figure*}
\centering
\includegraphics[width=\textwidth]{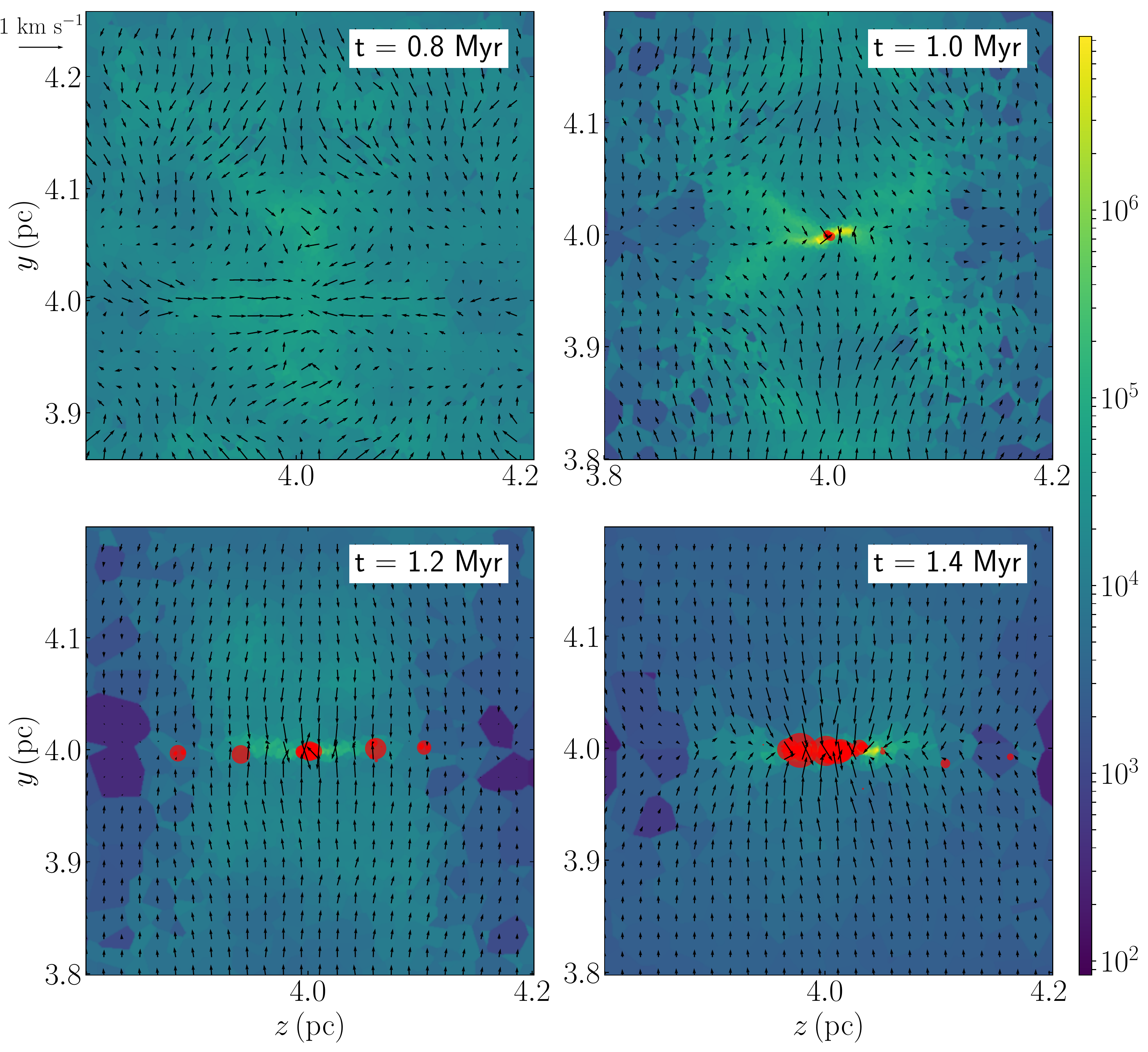}
\caption{
Same as Figure \ref{fig:zvx} but for the \mbox{COLA\_noB} model. The size of the domain is twice that of Figure \ref{fig:zvx}. The time spans from t=0.8 Myr to t=1.4 Myr.}\label{fig:zvxnob}
\end{figure*}

\begin{figure*}
\centering
\includegraphics[width=\textwidth]{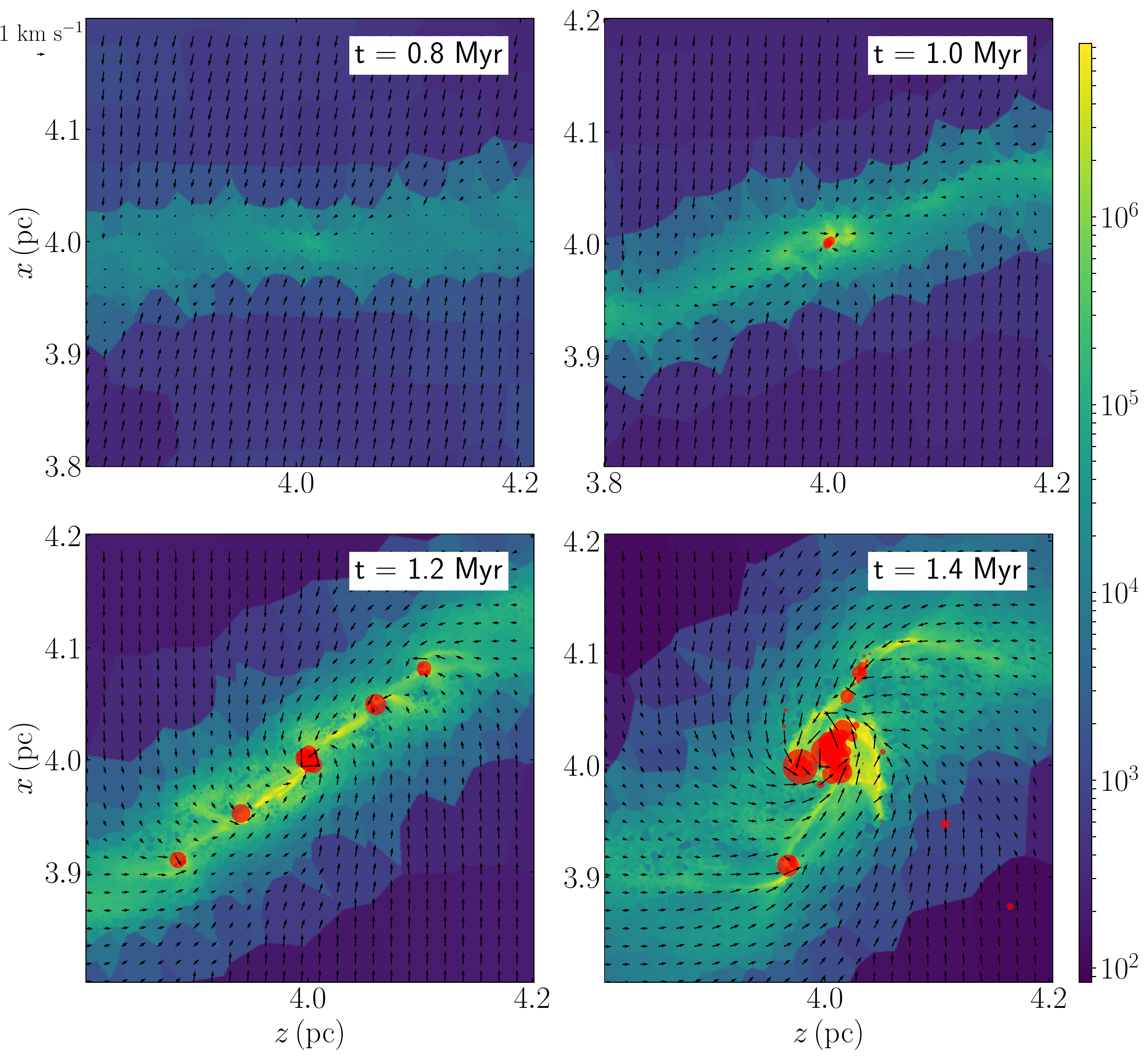}
\caption{
Same as Figure \ref{fig:zvy} but for the \mbox{COLA\_noB} model. The size of the domain is twice that of Figure \ref{fig:zvy}. The time spans from t=0.8 Myr to t=1.4 Myr.}\label{fig:zvynob}
\end{figure*}

\begin{figure*}
\centering
\includegraphics[width=\textwidth]{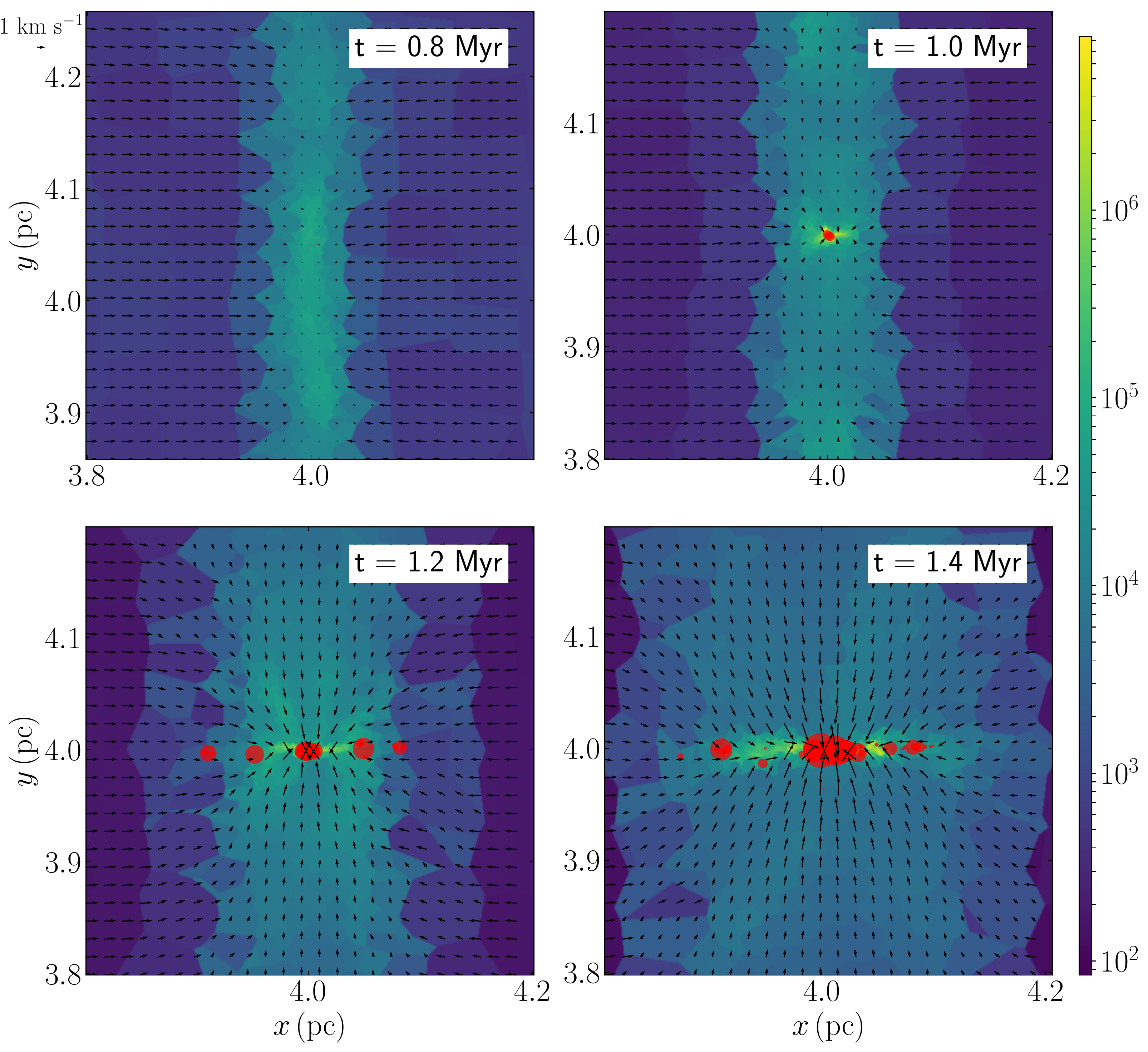}
\caption{
Same as Figure \ref{fig:zvz} but for the \mbox{COLA\_noB} model. The size of the domain is twice that of Figure \ref{fig:zvz}. The time spans from t=0.8 Myr to t=1.4 Myr.}\label{fig:zvznob}
\end{figure*}

For comparison, we run two more simulations with a uniform field (hereafter \mbox{COLA\_sameB}) and no field (hereafter \mbox{COLA\_noB}), respectively. All other parameters remain the same. Table \ref{tab:ic} lists the two models with their parameters. In \mbox{COLA\_sameB}, no sinks form (up to 3 Myr) because the gas density never gets high enough. It is not surprising that magnetic pressure hinders the formation of dense gas \citep[also see][]{2020ApJ...891..168W}. 

In \mbox{COLA\_noB}, however, sinks do form and show different behaviours compared to \mbox{MRCOLA}. Figures \ref{fig:zvxnob}, \ref{fig:zvynob}, \ref{fig:zvznob} show zoom-in slice plots for \mbox{COLA\_noB}. The zoom-in region is twice that in Figures \ref{fig:zvx}, \ref{fig:zvy}, \ref{fig:zvz}. First, sinks form earlier in \mbox{COLA\_noB} than \mbox{MRCOLA}. The first sink already forms at t=1.0 Myr in \mbox{COLA\_noB}. By t=1.4 Myr, there are 92 sinks present in the domain. This is more than the number of sinks (66) in \mbox{MRCOLA} at t=2.2 Myr which is 0.8 Myr later. Again, since we do not include feedback, we stop the \mbox{COLA\_noB} simulation at t=1.4 Myr. By this time, the most massive sink is 4.4 M$_\odot$.

Second, the overall star formation rate in \mbox{COLA\_noB} is higher than that in \mbox{MRCOLA}. Within 0.43 Myr, \mbox{COLA\_noB} sinks have a total mass of 41 M$_\odot$. The star formation rate is $9.5\times10^{-5}\epsilon$ M$_\odot$ yr$^{-1}$ where $\epsilon$ is the fraction of sink mass that is eventually converted to stars. In \mbox{MRCOLA}, within 0.72 Myr, 66 sinks form with a total mass of 33 M$_\odot$. The star formation rate is $4.6\times10^{-5}\epsilon$ M$_\odot$ yr$^{-1}$. Here we simply assume the same efficiency $\epsilon$ for both models. Then, \mbox{MRCOLA} has a star formation rate 2.1 times smaller than \mbox{COLA\_noB}. 

Third, \mbox{COLA\_noB} sinks form in a wider region compared to the fiducial model \mbox{MRCOLA}. Figure \ref{fig:zvynob} shows the y=4 pc slice from \mbox{COLA\_noB}. We can see that sinks spread over a region of $\sim$0.2 pc which is about four times the scale of the sink formation region in \mbox{MRCOLA}. At t=1.2 Myr, the sinks form along the dense gas elongation in multiple groups that are almost equally spaced. The elongation is the compression layer due to the collision. In \mbox{MRCOLA}, this elongation is squeezed by field loops into the central core, which is why we see a tighter cluster in Figure \ref{fig:zvy}. Figure \ref{fig:zvynob} also shows that the cluster follows the gas spiraling motion in the disk.

Compared to \mbox{MRCOLA}, \mbox{COLA\_noB} sinks are embedded in a better-defined dense gas disk. As shown in Figures \ref{fig:zvxnob} and \ref{fig:zvznob}, the sinks quickly settle in the y=4 pc plane after the first sink formation. Gas is falling from above and below the disk. This indicates a global collapse at t$\gtrsim$1.2 Myr that feeds the dense gas and cluster accretion in the disk. On the contrary, due to the complex magnetic fields, the global gas inflow in \mbox{MRCOLA} is only viable through those streamers, although a coherent inflow from above and below the cluster temporarily exists at t$\lesssim$1.6 Myr (see Figures \ref{fig:zvx} and \ref{fig:zvz}). The global collapse toward the central core is disturbed by the wrapping field, which is part of the reason (at large scales) that \mbox{MRCOLA} has a lower star formation rate than \mbox{COLA\_noB}. 

% The difference in star formation rate is somewhat contrary to our initial expectation. Simply based on the capability of dense gas production by CMR, K21 speculated that CMR would be able to accelerate star formation. However, as the comparison between \mbox{COLA\_noB} and \mbox{MRCOLA} shows here, the CMR-SF is indeed slower in the case with CMR. Meanwhile, CMR-SF also happens later, which is because CMR-SF has two phases (see \S\ref{subsec:fiducial}). The sink formation in phase-2 has to wait for the collapse of the filament formed during phase-1. It takes additional time to gather dense gas to a smaller region in \mbox{MRCOLA}. In the future, more explorations will show the impact on star formation rate from CMR.

\section{Discussion}\label{sec:discus}

\subsection{The role of CMR in cluster formation}\label{subsec:cmrrole}

Comparing \mbox{MRCOLA} and \mbox{COLA\_noB}, we can summarize two properties of CMR-SF that distinguish itself from other mechanisms. First, CMR-SF is confined in a relatively small region. The cluster is very tight at least during the accretion. Second, CMR-SF is relatively slow. Inflowing gas is only able to feed the cluster through streamers. The former is mainly due to the confinement of the helical field. The latter is again due to the helical field that is orthogonal to the gas inflow. 

In fact, if we re-think about the CMR process, it is essentially a process that gathers a large volume ($\sim$1 pc) of gas and compresses it to a small volume ($\sim$0.05 pc), creating an over-dense region that forms stars and also a potential well that accretes more gas. First, the colliding clouds bring gas from afar. The collision compresses the 3D spheres into a 2D sheet. Second, CMR compresses the gas into a dense filament. The reconnected field compresses the 2D sheet into a 1D filament. Third, the filament collapses into a dense core and a cluster forms. Gravity compresses the 1D filament into a 0D core. The three physical processes, i.e., the collision, the reconnection, and gravity, compress relatively diffuse gas into much denser gas through a step-by-step dimension reduction. 

As shown in \S\ref{subsec:control}, the cluster would be spread over a larger disk if the CMR mechanism is absent. One speculation is that the concentrated cluster in CMR is more likely to be bound, compared to the cluster without CMR. While we need more models to confirm the boundness, the difference makes CMR a possible explanation for those highly concentrated clusters in observations. However, as will be discussed in \S\ref{subsec:fb}, protostellar heating may suppress the fragmentation in the gas concentration in \mbox{MRCOLA}. Thus the number of stars is limited. If the first few stars accrete the majority of the gas, there may be more massive stars in the reduced cluster. Energetic feedback from the massive stars will eventually disperse the gas.

However, it is also possible that CMR-SF produces multiple clusters if the initial clouds are much larger. A collision between such clouds may form a much larger filament with multiple large fragments, each forming a star cluster. The large filament may or may not have the longitudinal collapse which pushes everything to the center. Future work will address this scenario.

The reconnected field from CMR makes it difficult for the gas inflow to feed the central star formation. In the absence of CMR, gas collpases and falls onto the star-forming disk easily through large-scale flows.  With CMR, gas falls to the center in similarly coherent flows initially. But a toroidal region around the central core narrows the angle of inflowing gas (almost no horizontal flow toward the core in Figures \ref{fig:zvx} and \ref{fig:zvz}). It is the magnetic pressure from the helical/toroidal field that hinders the gas inflow. Shortly after, the gas movement in the vicinity of the cluster becomes chaotic. The inflowing gas carry magnetic flux to the central region, changing the topology of the helical field. Everything becomes more chaotic and the gas inflow becomes inefficient. Now, gas can only reach the cluster through streamers. 

We can see that the helical field, which is the natural result from CMR, is responsible for the disturbance of the mass supply, which is why \mbox{MRCOLA} has a relatively low star formation rate. One thing that would be interesting to explore is the effect of magnetic field diffusion due to Ohmic resistivity and ambipolar diffusion. At such a small scale, the magnetic Reynolds number should become small, and magnetic diffusion should become important. If some amount of the magnetic energy is lost, the field will exert less pressure on the inflowing gas which may resume the coherent flow. In turn, the star formation rate may approach that in the case without CMR. Future work should address this uncertainty.

\subsection{Effect of protostellar feedback}\label{subsec:fb}

The concentrated cluster in \mbox{MRCOLA} is probably a result of the lack of protostellar heating. As we can see from Figure \ref{fig:tr}, the separation between the more massive sinks is $\lesssim$0.01 pc (2000 AU). Around each sink creation site, the typical gas density is $\gtrsim10^7$ cm$^{-3}$, corresponding to a Jeans scale of $\sim$0.005 pc at $\sim$20 K (the typical temperature in the CMR-filament). So the crowdedness is a result of fragmentation in the central dense core. 

However, protostars should form during the cluster formation because the free-fall time is just of order 10000 yr for a density of $10^7$ cm$^{-3}$. The protostellar accretion will inevitably heat the surroundings, thus increasing the overall Jeans scale in the core. Consequently, the number of fragmentations/sinks should be reduced. In fact, \citet{2009MNRAS.392.1363B} have studied the effect of protostellar radiative feedback. They found that the number of protostars was reduced by a factor of 4 in the radiation hydrodynamic simulation compared to the hydrodynamic simulation. Observationally, a recent ALMA result \citep{2017ApJ...837L..29H} showed that protostellar accretion can impact a volume of 2000 AU scale, which is larger than the sink separation in \mbox{MRCOLA}. Therefore, the sink number in our simulations is an upper-limit.

However, unless the feedback can stop the global collapse completely, the inflowing gas will keep transferring material to the central cluster, continuously feeding the protostellar accretion. Naively, we would expect more massive stars in \mbox{MRCOLA}. For instance, \citet{2011ApJ...740...74K} showed that the first generation of protostars from the initial fragmentation will keep accreting the inflowing gas, resulting in more massive stars. Similar results may happen in the CMR-SF if we consider protostellar heating. However, energetic feedback from the massive stars will probably halt further accretion and even completely disperse the gas. In the future, a CMR simulation with protostellar feedback will clarify the situation.

\subsection{Applicability of CMR to star-forming filaments}

The CMR-SF in the fiducial model occurs after the filament collapses. During the filament phase, there is no sink formation. The sterility of the filament, at least in the one model in this paper, raises the question the applicability of this model to star forming filaments in the Galaxy. 
For example, the Orion A filament is elongated and star formation is already ongoing in multiple (OMC-1/2/3/4) regions unlike our fiducial model where the filament first collapses into a core and then stars form in the core. Furthermore, how likely is the initial condition of antiparallel B-fields to occur in the cold ISM? The answers to these two questions will help clarify how common the CMR-SF mechanism is in filament and star formation.

To address the likelihood of antiparallel B-fields, we go back to the original proposal (K21) of the CMR mechanism. The K21 model, along with the fiducial model in this paper, was established specifically for the Stick filament in Orion A. At a first glance, the initial condition (see K21 figure 8) that led to CMR seemed unusual. However, it was what observational facts showed us. The field reversal around Orion A was clearly shown in \citet{1997ApJS..111..245H} and later in \citet{2019A&A...632A..68T}, using two different methods. The former showed that the field-reversal was a large-scale feature, not just a local small-scale stochastic fluctuation. Then, \citet{2019A&A...629A..96S} showed that the plane-of-the-sky B-field was nearly perpendicular to the filament. Combining the B-field observations and the two-component pattern in the PV-diagram (K21 figure 5), K21 set up the only possible initial condition in their figure 8, which surprisingly formed a filament at the collision front instead of a compression pancake. The K21 model successfully reproduced a number of observational facts, including the morphology (which was the motivation for the model as there were several ring/fork-like structures in the Stick), the density PDF, the line channel maps, and the PV-diagrams. Therefore, at least for the Stick filament, the CMR model was undoubtedly applicable.

In a broader context, how likely is the antiparallel B-field in the Milky Way and other galaxies? Are there molecular clouds formed via the CMR mechanism? The field-reversal is common in theoretical studies. In fact, with high enough resolution, a turbulent MHD simulation will show many field-reversal interfaces \citep[e.g.,][]{2018PhRvL.121p5101D,2020ApJ...895L..40C}. %In the cloud factory simulation series \citep{2020MNRAS.492.1594S}, there are field reversals in the simulated galactic disk. 
In the Galactic disk, Faraday Rotation measurements have long shown field reversal in our solar neighborhood, and several authors \citep[e.g.,][]{1983ApJ...265..722S,1994A&A...288..759H} have proposed a bisymmetric spiral disk field for the Milky Way. Such configurations have multiple field reversals along spiral arms in which a cloud-cloud collision would trigger CMR in a global simulation \citep[][note that we need a large dynamic range to be able to capture CMR]{2022ApJ...933...40K}. So, in both theoretical and observational senses, the CMR mechanism is a viable physical process that produces dense gas and clouds. Most recently, Faraday Rotation measurements showed that the Orion A cloud, the Perseus cloud, and the California cloud all sit between reversed B-fields \citep{2020IAUGA..30..103T}. It could be just a coincidence that all these clouds sat between two large-scale fields with inverted polarity. However, the CMR model showed that if there was a field-reversal and a cloud-cloud collision, the filament formation was automatically fulfilled at the field-reversing interface.

Strictly speaking, it is unlikely for the  B-fields to be exactly antiparallel in the sense of probability theory. In reality, there is also possibly small-scale fluctuation in the B-field orientation due to turbulence. K21 has briefly explored these effects. First, if the initial B-field had a relative angle of 20 degree on the two sides, the cloud-cloud collision was still able to create a dense filament (see their figure 26). But the filament in the middle of the compression pancake had a lower density and was shorter. With an initial B-field angle of 90 degree, the collision produced a diagonally symmetric dense patch (see their figure 27). In general, the trend was that CMR was less capable of forming a dense filament with a larger B-field tilting angle, which was not surprising because the reconnected fields were no longer loops in a flat plane. Second, K21 ran a CMR simulation with turbulence that was injected at the beginning. They found that the CMR with turbulence was still able to form the filament which had a smaller width and a wiggling morphology (see K21 figure 28). Theoretical studies have shown that turbulence accelerates magnetic reconnection \citep[e.g.,][]{1999ApJ...517..700L}. So, as long as the initial B-field is somewhat antiparallel, a cloud-cloud collision shall trigger CMR \citep[see detailed discussions in][]{2022ApJ...933...40K}. Here we use \textsc{Arepo} to explore CMR-SF with the initial B-field angle ranging from 10 to 90 degree at a step of 10 degree. We find that models with a tilting angle $\lesssim40$ degree are able to form sinks. Models with a tilting angle $\gtrsim50$ degree do not form sinks until the end of the simulation (3 Myr). Also, the larger the initial tilting angle, the less the sink formation, which is consistent with the trend of dense gas formation.

To answer the question about the sterility of the filament, we first need to discuss the fate of the Stick. As shown by the fiducial model in this paper, the filament will collapse along its main axis toward the center where a dense core forms. In the core, sink formation happens once the core density is significantly increased, and eventually a cluster emerges. In reality, will the Stick do the same thing? As shown by \citet{2000MNRAS.311...85F}, under the assumption of axisymmetry, an initially stable filament remains so against radial perturbation. The stability originates from the fact that the gravitational potential of the filament is independent of its radius, so its self-gravity will never dominate due to radial contraction. However, the same is not true for the longitudinal collapse, as the potential scales as $\sim L^{-1}$ \citep[$L$ is the filament length,][]{2000MNRAS.311...85F}, which indicates that the filament will inevitably collapse along its main axis. Currently, the Stick filament is still cold and starless. Most likely, the filament will collapse longitudinally and form stars.

Following the above reasoning, it becomes clear that the key to the sterility question is the length scale of the filament. The filament will collapse along its main axis eventually, so it cannot form stars before collapsing into the central core if it is too short and the gravitational instability does not have time to grow. The filament in the fiducial model has a length $\sim$1 pc, which is also the length scale of the Stick filament. In the simulation, the filament collapses within $\sim$1.4 Myr. As shown by \citet{2000MNRAS.311..105F}, the growth timescale for the gravity-driven mode is $\sim$1.8 Myr, longer than the collapse time of the filament \citep[also see][]{1997ApJ...480..681I}. Of course, the CMR-filament has rich sub-structures, some of which are quite dense ($\gtrsim 10^5$ cm$^{-3}$). They break the axisymmetry assumption in the \citet{2000MNRAS.311...85F} model. 

In fact, the dense sub-structures do not result from the traditional sense of filament fragmentation. They are created by magnetic tension and brought to the filament. Instead of forming a filament first and then letting it fragment, the CMR mechanism creates multiple clumpy sub-structures and then brings them together to constitute a filament (a bottom-up process). So the dense sub-structures exist from the beginning of the filament. They have a chance to grow denser if the filament lasts longer, possibly followed by star formation. We can imagine two 20 pc clouds colliding. Their sizes are $\sim$10 times larger than those in the fiducial model, and the collision timescale is also 10 times longer. Now it will take much longer for the filament to collapse into a central core. Sink formation should happen during the filament phase before the core formation. In fact, evidence has shown that the densest part (OMC-1) of the integral-shaped filament (ISF) in Orion A is undergoing a longitudinal collapse \citep{2017A&A...602L...2H}. It is also the region with the most active star formation in Orion A (the Trapezium cluster). Meanwhile, in the northern OMC-2/3 regions, star formation is also ongoing \citep[e.g.,][]{2021A&A...653A.117B}. For even longer filaments like Nessie \citep[$\gtrsim 100$  pc,][]{2010ApJ...719L.185J,2014ApJ...797...53G}, the longitudinal collapse may not bring everything into a central core before other dynamic processes breaking the filament, e.g., the Galactic shear and feedback. In fact, as shown by the mid infrared images \citep{2010ApJ...719L.185J}, the filament Nessie breaks into multiple dark sub-filaments, each showing signs of protostellar activity, indicating local collapses.

Alternatively, one can imagine the collision between two (almost) plane-parallel gas structures, whatever their physical and chemical states are (cold neutral medium vs. cold neutral medium, or warm neutral medium vs. warm neutral medium, or even cold neutral medium vs. warm neutral medium). As long as there are protruding structures on the surface that collide with (nearly) antiparallel B-fields, CMR shall be triggered and dense gas shall form \citep{2022ApJ...933...40K}. For instance, it can be the collision between the expanding bubble from a massive star or supernova and a wall of atomic/molecular gas. The surfaces of the bubble and the wall are likely not smooth but with ripples. As long as the bubble brings the antiparallel B-field, the collision shall trigger multiple CMR events at the collision interface. Each of these events will form a dense filament. Depending on the geometry, all the filaments may constitute a large filament or a web of filaments. Following the above reasoning about the filament collapse, we may see star formation happening at different locations. More interestingly, these star-forming clouds will have turbulence that originates from the chaotic reconnected field. The helical field will guide the incoming plasma into different directions, converting the coherent colliding velocity into chaotic turbulent energy, which gives a natural explanation of one origin of turbulence in molecular clouds. Future studies shall address all these physical processes. 

\section{Summary and Conclusion}\label{sec:conclu}

In this paper, we have investigated star formation in the context of collision-induced magnetic reconnection (CMR). Using the \textsc{Arepo} code, we have confirmed the filament formation via CMR, which was first shown in \citet{2021ApJ...906...80K}. With the sink formation module in \textsc{Arepo}, we have shown that the CMR-filament is able to not only form stars, but a mini star cluster. We stop the fiducial model simulation at t=2.2 Myr when there are 66 sinks in the computation domain. Further evolution of the gas and cluster is likely impacted by protostellar feedback, including outflows and radiative heating, which we currently do not consider.

At least in the fiducial model, the CMR star formation (CMR-SF) is a two-phase process. The first phase is the filament formation due to the magnetic reconnection. During this phase, we see a dense, clumpy filament with no star formation. The reason is that the cloud is not bound by gravity but by the surface pressure from the wrapping helical magnetic field. This starless phase lasts about 1.4 Myr. In the second phase, the filament starts to collapse longitudinally into a central dense core, shortly before the first sink formation in the core. With continuous fragmentation, a star cluster forms in the core. Those stars that become massive later form earlier and stay near the cluster center, while the outer part of the cluster preferentially consists of lower mass stars. The apparent mass segregation is indicative of competitive accretion.

Qualitatively, there are two distinctive features in CMR-SF. First, the number of clusters and their extent are limited. In our fiducial model, only one cluster forms and it is confined within a region of $\sim$0.05 pc. Both result from the highly concentrated dense gas, which is strongly confined by the helical/toroidal magnetic field and gravity. In comparison, the same model but without magnetic field has multiple cluster-forming sites that spread over a larger volume of $\gtrsim$0.2 pc. Second, because of the field, which acts like a surface shield, inflowing gas is only able to transfer material to the core/cluster through streamers. The limited gas inflow results in a relatively low star formation rate. Compared to the model without magnetic field, CMR-SF has an overall star formation rate a factor of 2 smaller.

In CMR-SF, the crowdedness of the cluster will probably result in more massive stars if protostellar feedback is included. For instance, the radiative heating will suppress fragmentation, thus limiting the number of stars in the cluster. So the same mass reservoir will supply more massive stars if they keep the accretion. Eventually, feedback from the massive stars will stop the accretion and disperse the gas.

\section{Acknowledgements}

An allocation of computer time from the UA Research Computing High Performance Computing (HPC) at the University of Arizona is gratefully acknowledged. RJS gratefully acknowledges an STFC Ernest Rutherford fellowship (grant ST/N00485X/1)  and HPC from the Durham DiRAC supercomputing facility (grants ST/P002293/1, ST/R002371/1, ST/S002502/1, and ST/R000832/1).

%%%%%%%%%%%%%%%%%%%%%%%%%%%%%%%%%%%%%%%%%%%%%%%%%%
\section*{Data Availability} 

The data underlying this article are generated from numerical simulations with the modified version of \textsc{Arepo} code. The data will be shared upon reasonable request to the corresponding author.

%%%%%%%%%%%%%%%%%%%% REFERENCES %%%%%%%%%%%%%%%%%%

% The best way to enter references is to use BibTeX:

\bibliographystyle{mnras}
\bibliography{ref} % if your bibtex file is called example.bib

% Alternatively you could enter them by hand, like this:
% This method is tedious and prone to error if you have lots of references
%\begin{thebibliography}{99}
%\bibitem[\protect\citeauthoryear{Author}{2012}]{Author2012}
%Author A.~N., 2013, Journal of Improbable Astronomy, 1, 1
%\bibitem[\protect\citeauthoryear{Others}{2013}]{Others2013}
%Others S., 2012, Journal of Interesting Stuff, 17, 198
%\end{thebibliography}

%%%%%%%%%%%%%%%%%%%%%%%%%%%%%%%%%%%%%%%%%%%%%%%%%%

%%%%%%%%%%%%%%%%% APPENDICES %%%%%%%%%%%%%%%%%%%%%

\appendix

\section{Overview of the fiducial model}\label{app:fiducial}

\begin{figure*}
\centering
\includegraphics[width=\textwidth]{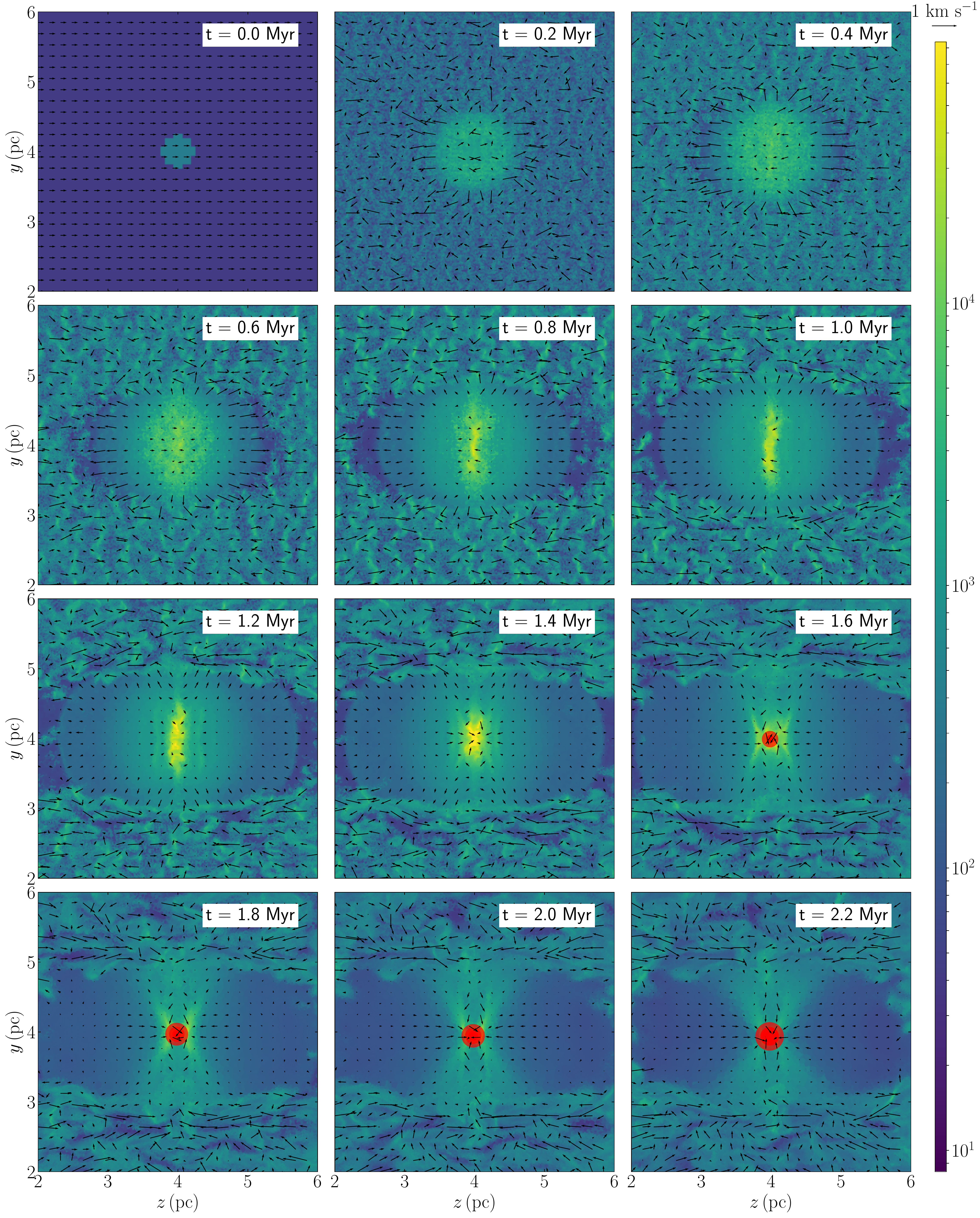}
\caption{
Density slice plots for the x=4 pc plane for model \mbox{MRCOLA}. The plane is the collision midplane between the two clouds. Each panel shows a snapshot of the simulation, with the time step labeled on the top-right of the panel. The computation domain spans from 0 to 8 pc on each side. We zoom in to the central 4 pc region from 2 pc to 6 pc. The color plot is in unit of $n_{\rm H_2}$ (cm$^{-3}$). The black arrows show the velocity vectors. Their lengths are proportional to the magnitudes. The red circles show the sink location. Their sizes are proportional to the sink mass.}\label{fig:mrcolavx}
\end{figure*}

\begin{figure*}
\centering
\includegraphics[width=\textwidth]{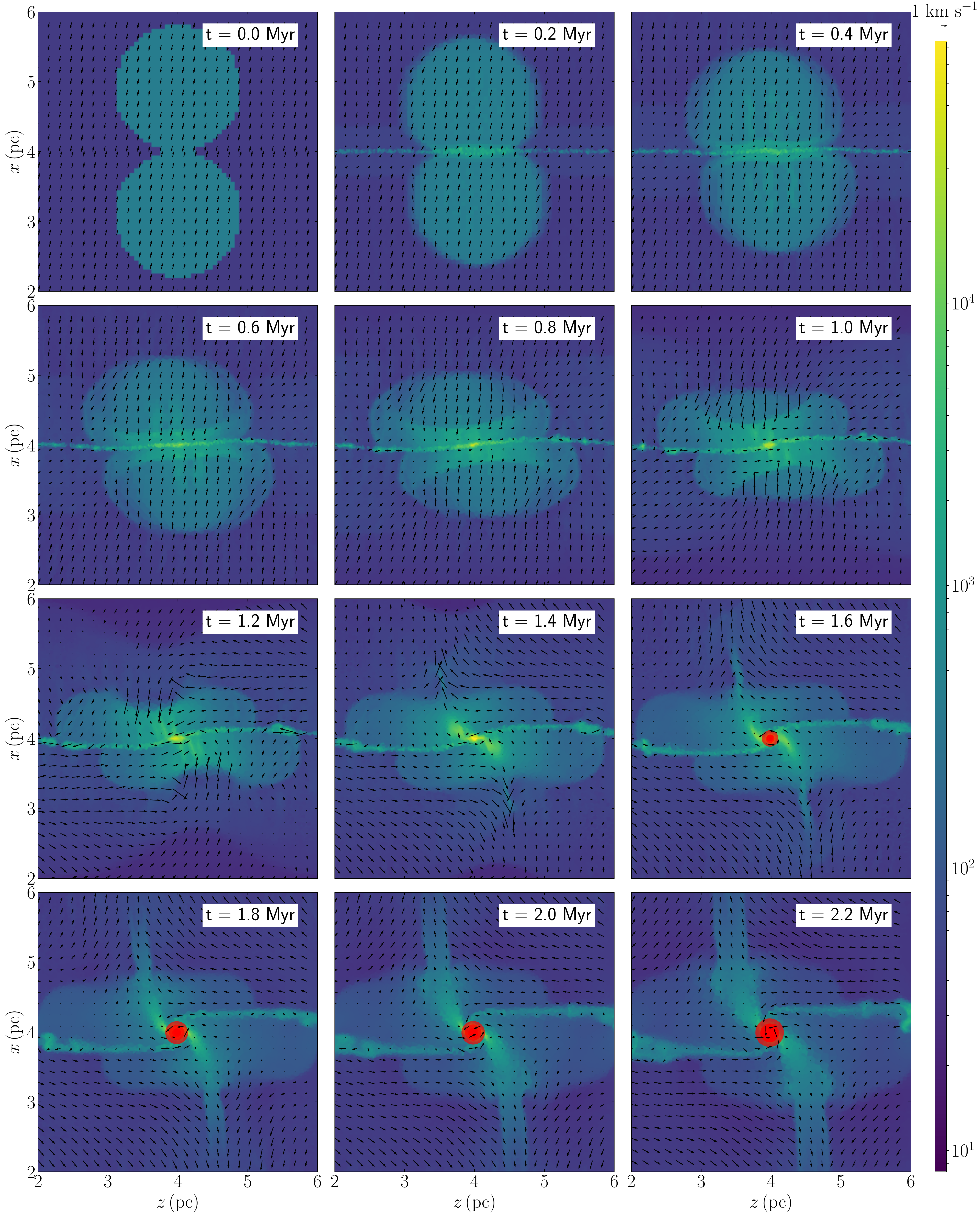}
\caption{
Same as Figure \ref{fig:mrcolavx}, but for the y=4 pc plane.}\label{fig:mrcolavy}
\end{figure*}

\begin{figure*}
\centering
\includegraphics[width=\textwidth]{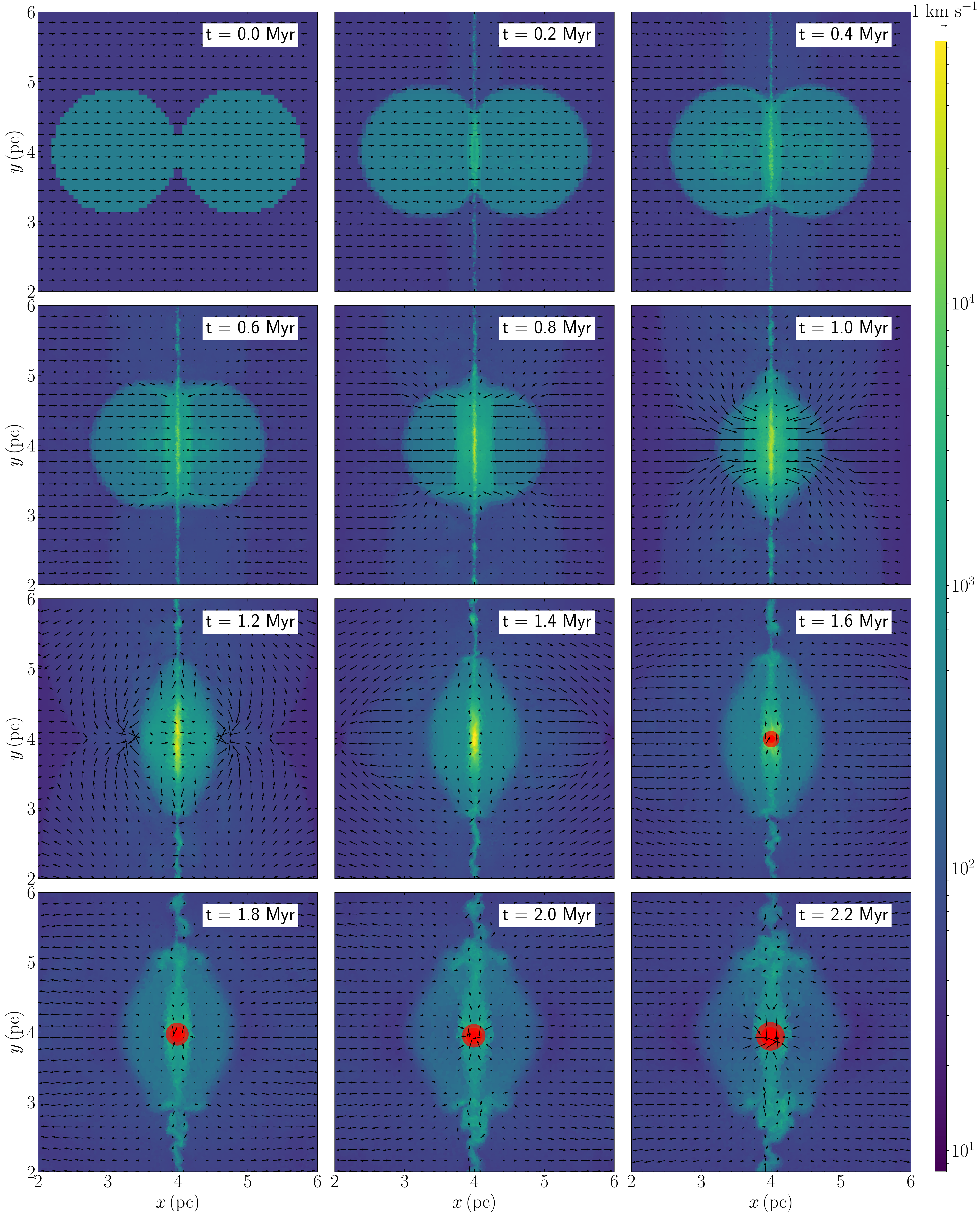}
\caption{
Same as Figure \ref{fig:mrcolavx}, but for the z=4 pc slices.}\label{fig:mrcolavz}
\end{figure*}

The results of the fiducial model are shown in Figures \ref{fig:mrcolavx}, \ref{fig:mrcolavy}, \ref{fig:mrcolavz}. The three figures show the gas density as a function of time for slices at x=4 pc, y=4 pc, and z=4 pc, respectively. The initial condition is shown in the first panel of each figure at t=0. 

Figure \ref{fig:mrcolavx} shows the slice at the collision midplane. This plane is the field reversal plane initially. At t=0.2 Myr, the two clouds collide and form a pancake in the plane. Note the ripples all over the plane. They are dense structures created by magnetic reconnection triggered by the collision, which was also seen in K21. At t=0.4 Myr, the wiggles grow thicker, and the pancake begins to shrink toward the central axis at z=4 pc. At t=0.6 Myr, the shrinking continues, and a dense filament forms along the central axis. The filament becomes more prominent at t=0.8 Myr and 1.0 Myr. The formation of the filament, instead of a dense pancake, indicates that CMR happens.

In Figure \ref{fig:mrcolavx}, after t=1.2 Myr, the filament begins to collapse along the main axis. At t=1.6 Myr, the collapsing gas converges to form a dense core in the center. The core persists until the end of the simulation. A sink particle forms at t=1.48 Myr near the center of the filament. The time is when the filament starts to collapse longitudinally. \mbox{MRCOLA} shows that CMR can form stars.

Starting from t=1.4 Myr, there is an X-shaped outflow between the horizontal and vertical collapse (also see velocity vectors in Figure \ref{fig:mrcolavxlin}). We can see that part of the inflowing gas is carried away by the X-outflow. From the previous analysis of Figure \ref{fig:mrcolavy} we know that material spirals into the filament in the y=4 pc plane. But the X-outflow indicates that the horizontal accretion is limited within a narrow y-range. 

At t=2.2 Myr, Figure \ref{fig:mrcolavxlin} shows that the X-outflow is less prominent while the horizontal and vertical collapses continue. Meanwhile, Figure \ref{fig:mrcolavz} shows that in the z=4 pc plane the dense core and the cluster accrete from all directions. As shown in Figure \ref{fig:mrcolavxlin}, all the sinks concentrate near the core center.

Figure \ref{fig:mrcolavy} shows the y=4 pc slices for \mbox{MRCOLA}. Here we see the cross-section of the filament which is at the center of the plots. The velocity vectors indicate rotation in the z-x plane. At t$\gtrsim$1.2 Myr, the gas spirals in to the filament. 

Figure \ref{fig:mrcolavz} shows the z=4 pc slices for \mbox{MRCOLA}. Here we have the side view of the two colliding clouds. One thing to note is the density waves created by the collision. They propagate to the x=0 and x=8 pc boundaries and will enter the domain again due to the periodic boundary condition. However, for our simulation time, they do not affect the central filament. The CMR process and sink formation are not impacted.

%%%%%%%%%%%%%%%%%%%%%%%%%%%%%%%%%%%%%%%%%%%%%%%%%%

% Don't change these lines
\bsp	% typesetting comment
\label{lastpage}
\end{document}